%% file: arxiv_version.tex
\documentclass[11pt]{article}
\usepackage{fullpage}
\usepackage[T1]{fontenc}
\usepackage[utf8]{inputenc}
\usepackage[colorlinks=true,allcolors=blue]{hyperref}
\usepackage{amsmath, amssymb, amsthm, amsfonts, graphicx,color,subcaption, enumerate, bm, cleveref, enumitem, array}

\usepackage{multicol, multirow}
\usepackage{thmtools, thm-restate}
\usepackage{ragged2e}
\usepackage{lineno}
\usepackage{subfiles}
\graphicspath{{graphics/}}

\newtheorem{theorem}{Theorem}[section]
\newtheorem{lemma}[theorem]{Lemma}
\newtheorem{corollary}[theorem]{Corollary}

\newtheorem{claim}[theorem]{Claim}
\newtheorem{observation}[theorem]{Observation}
\newtheorem{definition}[theorem]{Definition}

\newcommand{\LOCAL}{\mathsf{LOCAL}}
\newcommand{\CONGEST}{\mathsf{CONGEST}}
\def\polylog{\operatorname{polylog}}
\newcommand{\eps}{\varepsilon}
\renewcommand{\epsilon}{\varepsilon}
\newcommand{\poly}{\operatorname{poly}}

\newcommand{\R}{\mathbb{R}}

\newcommand{\cM}{\mathcal{M}}

\newcommand{\cR}{\mathcal{R}}
\newcommand{\cT}{\mathcal{T}}
\newcommand{\cS}{\mathcal{S}}

\newcommand{\cG}{\mathcal{G}}
\newcommand{\OO}{\widetilde{O}}
\newcommand{\VD}{\textsc{VD}}

\newcommand{\cc}{\textsc{cc}}

\newcommand{\PD}{\textsc{PD}}
\newcommand{\tG}{\widetilde{G}}
\newcommand{\oH}{\overline{H}}
\newcommand{\ow}{\overline{w}}
\DeclareMathOperator{\dist}{dist}
\renewcommand{\emptyset}{\varnothing}

\title{Fully Scalable Massively Parallel Algorithms \\ for Embedded Planar Graphs}

\author{Yi-Jun Chang \\ National University of Singapore \and Da Wei Zheng \\ University of Illinois Urbana-Champaign}

\date{}

\begin{document}

\maketitle
\thispagestyle{empty}

\subfile{abstract}

\pagebreak

\tableofcontents
\thispagestyle{empty}

\pagebreak
\clearpage
\pagenumbering{arabic} 
\subfile{main}

\end{document}

%% file: abstract.tex
\begin{abstract}
We consider the \emph{massively parallel computation} (MPC) model, which is a theoretical abstraction of large-scale parallel processing models such as MapReduce. 
In this model, assuming the widely believed  1-vs-2-cycles conjecture, solving many basic graph problems in $O(1)$ rounds with a strongly sublinear memory size per machine is impossible. 
We improve on the recent work of Holm and Tětek [SODA 2023] that bypass this barrier for problems when a planar embedding of the graph is given.
In the previous work, on graphs of size $n$ with $O(n/\mathcal{S})$ machines, the memory size per machine needs to be at least $\mathcal{S} = n^{2/3+\Omega(1)}$, whereas we extend their work to the \emph{fully scalable} regime, where the memory size per machine can be $\mathcal{S} = n^{\delta}$ for any constant $0< \delta < 1$.
We thus give the first constant round fully scalable algorithms for embedded planar graphs for 
the problems of
(i) connectivity
and
(ii) minimum spanning tree (MST).

Moreover, we show that the $\varepsilon$-emulator of Chang, Krauthgamer, and Tan [STOC 2022] can be incorporated into our recursive framework to obtain constant-round $(1+\varepsilon)$-approximation algorithms for the problems of computing 
(iii) single source shortest path (SSSP),
(iv) global min-cut,
and
(v) $st$-max flow.
All previous results on cuts and flows required linear memory in the MPC model.
Furthermore, our results give new algorithms for problems that implicitly involve embedded planar graphs.
We give as corollaries of our result the constant round fully scalable algorithms for
(vi) 2D Euclidean MST using $O(n)$ total memory 
and
(vii) $(1+\varepsilon)$-approximate weighted edit distance using $\widetilde{O}(n^{2-\delta})$ memory.

Our main technique is a recursive framework combined with novel graph drawing algorithms that allow us to compute smaller embedded planar graphs in constant rounds in the fully scalable setting.
\end{abstract}

%% file: main.tex
\section{Introduction}
We consider the \emph{massively parallel computation} (MPC) model introduced by Karloff, Suri, and Vassilvitskii~\cite{karloff2010MapReduce}, which is a theoretical abstraction of large-scale parallel processing models such as MapReduce~\cite{dg04}. In comparison to the classical PRAM model of parallel computing, the MPC model allows a substantial amount of local computation in each round, making it a more realistic model for practical parallel computation. 
The MPC model has received much attention in recent years~\cite{andoni2014parallel, Andoni2018, behnezhad2019near, balliu2023optimal, coy2022deterministic, czumaj2021simple, chang2019complexity, czumaj2018round, dhulipala2020parallel, EpastoMMZ22, ghaffari2018improved, ghaffari2019conditional, im2017efficient}.

In the MPC model, the input is initially partitioned into machines with a memory of $\Theta(\cS)$ words. The total memory size is linear in the size of the input. For example, if the input is a graph with $n$ vertices and $m$ edges and $n=\Theta(m)$, then each word in the memory stores $O(\log n)$ bits and the total number of machines is $O\left( \frac{m}{\cS}\right)$.
The machines communicate with each other in synchronous rounds.
In each round, each machine can send and receive $O(\cS)$ messages of $O(\log n)$ bits.
After communicating with other machines, each machine can perform a local computation of $\poly(\cS)$ time. The main complexity measure is the number of rounds needed to solve the problem under consideration. 

Our focus in this paper is on the \emph{fully scalable} regime, where the local memory size can be an arbitrarily small polynomial, i.e., $\cS = n^\delta$ for any constant $\delta > 0$. 
Designing algorithms for this regime is considerably more difficult than the setting where $\cS$ is a \emph{fixed} polynomial of $n$, and it has been the explicit goal of many recent papers, including~\cite{Andoni2018, AndoniSZ19, coy2022deterministic, czumaj2021simple, EpastoMMZ22, GhaffariU19}. 
%In the era of big data, it is increasingly important to keep the round complexity constant while allowing for an arbitrarily small constant $\delta$.

This paper considers planar graphs with a given embedding.
We assume that each vertex is assigned a coordinate in the plane, and edges are represented by straight lines without crossings, although our approach can also accommodate edges with few crossings and more complex families of curves.
This assumption is commonly observed in various applications where embeddings are readily available.
For instance, real-world map data includes GPS coordinates and descriptions of roads.
Often, the only way we know that a graph is planar is when we have an explicit planar embedding of the graph on the plane
%the only reason we know that a graph is planar is because we have an explicit planar embedding of the graph in the plane.

%There are no unconditional lower bounds on round complexity without implying a strong circuit lower bound~\cite{roughgarden2018shuffles}.  
%As such, \emph{conditional lower bounds} are required to demonstrate hardness of problems in the MPC model.
In sequential models of computation, it usually does not matter if we are given an embedding or not, as there are known algorithms (e.g.~\cite{schnyder1989planar}) to find a straight-line embedding in linear time.
Surprisingly, in the MPC model, there are conditional lower bounds that indicate a separation between whether or not we have an embedding.
%One successful tool to establish conditional lower bounds is the widely believed \emph{1-vs-2-cycles}
The widely believed \emph{1-vs-2-cycles conjecture}~\cite{Beame13,kiveris2014connected,karloff2010MapReduce,rastogi2013finding,roughgarden2018shuffles}, states that distinguishing between an $n$-vertex cycle and two $(n/2)$-vertex cycles (both planar graphs) requires $\Omega(\log n)$ rounds in the MPC model if the local memory size $\cS$ is at most $n^\delta$ for any $0<\delta< 1$.
Assuming this conjecture, many
basic graph problems, including minimum spanning tree (MST) and counting connected components, cannot be solved in constant rounds in the 
MPC model in the fully scalable regime, even for planar graphs.

Such a barrier can be bypassed for some geometric problems. In particular,  Andoni, Nikolov, Onak, and Yaroslavtsev~\cite{andoni2014parallel} showed that for any constant $d$ and $\epsilon$, a $(1+\epsilon)$-approximate solution for the Euclidean MST problem on $n$ points in $\R^d$ can be computed in $O(1)$ rounds in the MPC model in the fully scalable regime.

In contrast, for general graphs, even for the problem of computing connected components, the state-of-the-art algorithm in the MPC model in the fully scalable regime requires $O(\log D) + O(\log_{m/n} \log n)$ rounds~\cite{behnezhad2019near}, where $D$ is the diameter of the input graph. It is unlikely that this bound can be significantly improved due to the $\Omega(\log D)$ conditional lower bound based on the 1-vs-2-cycles conjecture~\cite{behnezhad2019near}.

On the other hand, for planar graphs with a given straight-line embedding, the recent work of Holm and T\v{e}tek~\cite{HolmT23} obtained constant-round  MPC algorithms for connected components, minimum spanning tree (MST), and $O(1)$-approximation of $st$-shortest path, diameter, and radius for the case where the local memory size is $\cS = n^{2/3 + \Omega(1)}$. 
This work showed that it is possible to bypass the 1-vs-2-cycle conjecture if we had an explicit planar embedding of a graph and had $n^{2/3+\Omega(1)}$ space per machine.
Their work left one major open question:

%\bigskip
\begin{center}
\begin{minipage}{0.85\textwidth}
\textit{Do there exist $O(1)$-round MPC algorithms for embedded planar graphs in the fully scalable regime where the local memory size is $\cS = n^{\delta}$ for any constant $\delta > 0$?}
\end{minipage}
\end{center}
%\bigskip

We answer this question in the affirmative by presenting a new framework for embedded planar graphs that solves a large class of problems in the fully scalable regime. 
Using this framework, we give the first constant-round algorithm for connected components and MST in this regime.
Furthermore, we are able to improve the approximation factor to $(1+\eps)$ for $st$-shortest path, and obtain $(1+\eps)$ approximations for more challenging fundamental graph problems such as single source shortest path (SSSP), all-pairs shortest path (APSP), max-flow, and min-cut.
Prior work on $(1+\eps)$-approximate distances, cuts, and flows required at least linear local memory $\cS = \Omega(n)$.
Our work presents the first constant-round algorithm for these fundamental problems in this more challenging memory regime.

Our main idea is to leverage the power of recursion by reducing a problem on an embedded planar graph into a problem on a smaller embedded planar graph.
We use a combination of \emph{geometric cuttings} and \emph{$r$-divisions} to partition the graph into regions with few edges crossing the boundary of each region. 
Our key new ingredient is a novel two-round nested recursive framework where we first use one outer round of recursion to reduce the graph size in each region. 
We then use a second layer of recursion on the graph we obtain by \emph{gluing} the smaller graphs from each region together.
In order to facilitate the recursion, our framework employs graph drawing algorithms to compute explicit embeddings of graphs when the graphs are small enough, and to glue graphs together in each (outer) recursive step.

\subsection{Our contributions}\label{sec:results}
\begin{table}[ht]
    \setlength\extrarowheight{3pt}
    \centering
    \begin{tabular}{ccccc}
        \multicolumn{2}{c}{Problem} & Total space & \shortstack{Memory \\per machine} & Source\\ \hline
       \multirow{9}{*}{\shortstack{Embedded \\ planar \\ graphs}} 
       & \multirow{2}{*}{Connected Components} & $O(n)$ & $n^{2/3+\Omega(1)}$ & \cite{HolmT23} \\
        & & $O(n)$ & $n^{\delta}$ & \Cref{thm:cc} \\ \cline{2-5}
        & \multirow{2}{*}{Minimum Spanning Tree} & $O(n)$ & $n^{2/3+\Omega(1)}$ & \cite{HolmT23}\\
        & & $O(n)$ & $n^{\delta}$ & \Cref{thm:mst} \\ \cline{2-5}
        &     $O(1)$-approx. SSSP & $O(n)$ & $n^{2/3+\Omega(1)}$ & \cite{HolmT23}\\
        & $(1+\eps)$-approx. SSSP & $O(n)$ & $n^{\delta}$ & \Cref{thm:sssp} \\ \cline{2-5}
        & $(1+\eps)$-approx. APSP & $O(n^2)$ & $n^{\delta}$ & \Cref{thm:apsp} \\ \cline{2-5}
        & $(1+\eps)$-approx. global min cut & $O(n)$ &  $n^{\delta}$ & \Cref{thm:global_mincut}\\ \cline{2-5}
        & $(1+\eps)$-approx. $st$-max flow & $O(n)$ &  $n^{\delta}$ & \Cref{thm:maxflow} \\ \hline
        
       \multirow{3}{*}{\shortstack{2D\\ Euclidean\\ MST}} & $(1+\eps)$-approx. & $O(n)$ & $n^{\delta}$ & \cite{andoni2014parallel} \\
       & Exact & $O(n)$ & $n^{2/3+\Omega(1)}$ & \cite{HolmT23} \\
       & Exact & $O(n)$ & $n^{\delta}$ &  \Cref{cor:EMST} \\ \hline
        
        \multirow{3}{*}{\shortstack{Edit \\ Distance}}
        & $(3+\eps)$-approx. & $\OO(n^{(9-4\delta)/5})$ & $n^{\delta}$ & \cite{BoroujeniGS21} \\
        & $(1+\eps)$-approx. & $\OO(n^{2-\delta})$ & $n^{\delta}$ & \cite{HajiaghayiSS19} \\
        & $(1+\eps)$-approx. weighted & $\OO(n^{2-\delta})$ & $n^{\delta}$ &\Cref{thm:edit_dist} \\\hline
    \end{tabular}
    \caption{Highlights of this work in comparison to prior work. SSSP stands for single source shortest paths, and APSP stands for all pairs shortest paths. All entries take $O(1)$ rounds.}
    \label{tab:results}
\end{table}

The approach of Holm and T\v{e}tek~\cite{HolmT23} is only applicable to machines with local memory $\cS = n^{2/3 + \Omega(1)}$. This lower bound is due to the fact that the computation of the $r$-division has to be done in one machine, and there is a tradeoff between the local memory size and the total number of boundary vertices. The bound $\cS = n^{2/3 + \Omega(1)}$ is the result of balancing the two quantities. It was asked in~\cite[Open question 5]{HolmT23} whether it is possible to extend their framework to the case where $\cS = n^{2/3 - \Omega(1)}$. 
Our first contribution is to resolve this problem by developing a new recursive framework, which allows us to extend the results in~\cite{HolmT23} to the fully scalable regime where we are allowed to set $\cS = n^{\delta}$ for any constant $\delta > 0$.  
Furthermore, our recursive framework can be made completely deterministic, while the algorithms of Holm and T\v{e}tek were randomized.

 \begin{restatable}[Connected Components]{theorem}{thmCC}\label{thm:cc}
There is an algorithm that returns the number of connected components of 
an embedded planar graph $G$ with $n$ vertices
in $O(1)$ rounds 
using $\Theta(\cS)$ space per machine and $O(n/\cS)$ machines where $\cS = n^{\delta}$ for any constant $\delta > 0$.
\end{restatable}

 \begin{restatable}[Minimum Spanning Forest]{theorem}{thmMST}\label{thm:mst}
    There is an algorithm that returns a minimum spanning forest of an embedded planar graph $G$ with $n$ vertices
    in $O(1)$ rounds 
    %in expectation and with high probability, 
    using $\Theta(\cS)$ space per machine and $O(n/\cS)$ machines where $\cS = n^{\delta}$ for any constant $\delta > 0$.
\end{restatable}

Since Delaunay triangulations can be computed in $O(1)$ rounds~\cite{Goodrich97}, we obtain the first constant-round Euclidean MST algorithm for $\R^2$ that works in the fully scalable regime.

\begin{corollary}[Euclidean MST]
\label{cor:EMST}
There is an algorithm that computes the Euclidean MST of 
a set $P$ of $n$ points in $\R^2$
in $O(1)$ rounds 
%in expectation and with high probability, 
using $\Theta(\cS)$ space per machine and $O(n/\cS)$ machines where $\cS = n^{\delta}$ for any constant $\delta > 0$.
\end{corollary}

\paragraph{Shortest paths.} Our second contribution is to show that our framework can be combined with the recent $\eps$-emulator of Chang, Krauthgamer, and Tan~\cite{ChangKT22}, and this allows us to design $O(1)$-round MPC algorithms for $(1+\eps)$-approximation, for any constant $\eps > 0$, of \emph{single-source shortest paths} (SSSP) and \emph{shortest cycle} with local memory size $\cS = n^{\delta}$, for any constant $\delta > 0$. These results improve the distance computation algorithms in~\cite{HolmT23}, which only achieve an approximation ratio of a fixed constant. This result resolves~\cite[Open question 6]{HolmT23}, which asks whether a better distance approximation is possible. 

\begin{restatable}[$(1+\eps)$-approximate Shortest Cycle]{theorem}{thmshortestcycle}\label{thm:shortest_cycle}
    There is an algorithm that computes the length of a $(1+\eps)$-approximate shortest cycle 
    of an embedded planar graph $G$ with $n$ vertices 
    in $O(1)$ rounds 
    %in expectation and with high probability, 
    using $\Theta(\cS)$ space per machine and $O(n/\cS)$ machines where $\cS = n^{\delta}$ for any constant $\delta > 0$.
\end{restatable}

\begin{restatable}[$(1+\eps)$-approximate SSSP]{theorem}{thmsssp}\label{thm:sssp}
    There is an algorithm that computes $(1+\eps)$-approximate single source shortest paths 
    of an embedded planar graph $G$ with $n$ vertices 
    in $O(1)$ rounds 
    %in expectation and with high probability, 
    using $\Theta(\cS)$ space per machine and $O(n/\cS)$ machines where $\cS = n^{\delta}$ for any constant $\delta > 0$.
\end{restatable}

As a corollary, we immediately obtain a constant-round 
 MPC algorithm for $(2+\eps)$-approximation for both radius and diameter in the fully scalable regime, as it is well-known that the longest shortest path distance from any given source vertex gives a $2$-approximation of both radius and diameter. 
\begin{corollary} [$(2+\eps)$-approximate diameter and radius]
    There is an algorithm that computes $(2+\eps)$-approximate diameter and radius
    of an embedded planar graph $G$ with $n$ vertices 
    in $O(1)$ rounds 
    %in expectation and with high probability, 
    using $\Theta(\cS)$ space per machine and $O(n/\cS)$ machines where $\cS = n^{\delta}$ for any constant $\delta > 0$.
\end{corollary}

Using the same ideas and $O(n^{2})$ total space, we can solve the $(1+\eps)$-approximate all-pairs shortest paths (APSP) problem. While this uses significantly more space than the size of the graph, $\Omega(n^2)$ total space is required to output the answer.
\begin{restatable}[$(1+\eps)$-approximate APSP]{theorem}{thmapsp}
\label{thm:apsp}
    There is an algorithm that computes a $(1+\eps)$-approximate shortest path 
    for all pairs of vertices
    of an embedded planar graph $G$ with $n$ vertices 
    in $O(1)$ rounds 
    %in expectation and with high probability, 
    using $\Theta(\cS)$ space per machine and $O(n^2/\cS)$ machines where $\cS = n^{\delta}$ for any constant $\delta > 0$.
\end{restatable}

As a corollary, this gives a method for finding  $(1+\eps)$-approximate diameter and radius, albeit using quadratic instead of linear total memory.
%The diameter is the maximum shortest path distance which we can easily find the maximum of in a $O(n^2)$ table in $O(1)$ rounds in the MPC model.
%The radius is the minimum maximum shortest path distance for every vertex, we can compute the maximum distance for each vertex in parallel in $O(1)$ rounds, then select the minima across the vertices in another $O(1)$ rounds.
\begin{corollary} [$(1+\eps)$-approximate diameter and radius]
    There is an algorithm that computes a $(1+\eps)$-approximate diameter and radius
    of an embedded planar graph $G$ with $n$ vertices 
    in $O(1)$ rounds 
    %in expectation and with high probability, 
    using $\Theta(\cS)$ space per machine and $O(n^2/\cS)$ machines where $\cS = n^{\delta}$ for any constant $\delta > 0$.
\end{corollary}

\paragraph{Planar duals.} Our third contribution is to show that a graph that contains as a minor the \emph{dual graph} of the given embedded planar graph can be constructed in $O(1)$ rounds in the fully scalable regime. The total space needed is $O(n)$. This dual graph construction, together with our shortest cycle algorithm, implies that a $(1+\eps)$-approximation of \emph{minimum cut} and \emph{maximum flow} can also be computed in $O(1)$ rounds in the fully scalable regime. This result resolves~\cite[Open question 3]{HolmT23}, which asks for an MPC algorithm for the minimum cut problem in embedded planar graphs.

\begin{restatable}[$(1+\eps)$-approximate global min-cut]{theorem}{thmmincut}\label{thm:global_mincut}
    There is an algorithm to compute 
    the global min-cut 
    of an embedded planar graph $G$ with $n$ vertices 
    in $O(1)$ rounds 
    %in expectation and with high probability, 
    using $\Theta(\cS)$ space per machine and $O(n/\cS)$ machines where $\cS = n^{\delta}$ for any constant $\delta > 0$.
\end{restatable}

\begin{restatable}[$(1+\eps)$-approximate $st$-max-flow]{theorem}{thmmaxflow}\label{thm:maxflow}
    There is an algorithm to compute a $(1+\eps)$-approximate maximum flow between 
    two vertices of any embedded planar graph $G$ with $n$ vertices 
    in $O(1)$ rounds 
    %in expectation and with high probability, 
    using $\Theta(\cS)$ space per machine and $O(n/\cS)$ machines where $\cS = n^{\delta}$ for any constant $\delta > 0$.
\end{restatable}

The computations of distances, cuts, and flows are difficult problems in the MPC model in that all existing works on this topic require a local memory of at least linear size: $\cS = \Omega(n)$~\cite{becker2021near,henzinger2019deterministic,ghaffari2018congested,ghaffari2020massively}. Our work is the first one that solves these problems in the fully scalable regime.

The list of problems presented here is not exhaustive, as our recursive framework is very versatile.
With this framework, we can find algorithms 
for variants of the problems that we discuss (e.g., computing labels for connected components, recovering shortest paths, flows, and cuts)
and for problems beyond the ones presented (e.g., verifying that the embedding is planar and finding a bipartition of the graph).

\paragraph{Work efficiency.}
Although our work mainly focuses on solving problems in the MPC model with constant round complexity, we emphasize that all our algorithms are extremely \emph{work-efficient}. The randomized variants of our algorithms perform nearly linear total work of $O(n\log n)$ for an input $n$-vertex embedded planar graph $G$, except for $(1+\eps)$-approximate APSP and its related problems, which require quadratic work.
The deterministic variants of our algorithm only require slightly more work, $O(n^{1+\alpha})$ for any constant $\alpha > 0$.

\paragraph{An application to edit distance.} 
Surprisingly, we are able to show that our techniques for planar graphs have implications 
for string algorithms.
Our new SSSP algorithm implies a new $(1+\eps)$-approximate edit distance algorithm in the MPC model 
whose performance matches the state of the art~\cite{HajiaghayiSS19}.
The computation of edit distance is one of the most fundamental problems in pattern matching, due to its wide range of applications in computational biology, natural language processing, and information theory. 

The first MPC algorithm for approximate edit distance was given in~\cite{boroujeni2021approximating}, which ran in $O(\log n)$ rounds and has an approximation ratio of $3+\eps$.  The subsequent work~\cite{HajiaghayiSS19} showed that an $(1+\eps)$ approximation of the edit distance between two strings of length at most $n$ can be computed in $O(1)$ rounds in the MPC model with $\widetilde{O}(n^\delta)$ memory per machine and $O(n^{2-2\delta})$ machines.

By a reduction to planar SSSP, we present a new edit distance algorithm in the MPC model matching the result obtained in~\cite{HajiaghayiSS19}.  Moreover, our result gives the first 
 MPC algorithm for \emph{weighted} version of the  $(1+\eps)$-approximate edit distance. Prior work~\cite{HajiaghayiSS19} heavily relied on the cost of insertion, deletion, and substitutions being equal. Our algorithm works for the weighted version of the edit distance problem, as long as the weight function is symmetric in the sense that the edit distance from any string $s$ to any string $t$ is the same as the edit distance from $t$ to $s$. 

\begin{restatable}[Edit distance]{theorem}{thmeditdist}\label{thm:edit_dist}
There exists an algorithm that 
given as input two strings $s$ and $t$ of length $n$ and a symmetric weight function $w$, 
computes a $(1+\eps)$-approximate edit distance between $s$ and $t$
in $O(1)$ rounds doing $\OO(n^2)$ work using $\OO(S)$ space per machine
and $\OO(n^2/S^2)$ machines where $S = n^{\delta}$ for any constant $\delta > 0$.
\end{restatable}

\subsection{Technical overview}\label{sec:technical}

\paragraph{Technical barriers.} As discussed earlier, the reason that the approach of Holm and T\v{e}tek~\cite{HolmT23} is only applicable to machines with local memory $\cS = n^{2/3 + \Omega(1)}$ is that they need to make sure that the total number of boundary vertices is at most $\cS$, as the size of the compressed graph is linear in this number and they need to process this graph in one machine.
We begin by explaining the technical barriers that their algorithm runs into when $\cS = n^{\delta}$.
The first is that with less memory, the decomposition that we compute and distribute to all machines needs to be much smaller, around $n^{\delta}$ in size.
This means that the number of boundary vertices is much larger, on the order of $n^{1-O(\delta)}$, so we cannot hope to store all the boundary vertices on one machine.
This is not immediately an issue, as it may be possible to use recursion to repeatedly decompose the graph into smaller pieces until they are small enough to solve on one machine,
%. 
%We cannot fully solve the problem for any boundary vertex, 
so we need to return some sort of compressed representation of the graph whose size is proportional to the number of boundary vertices, put the graph together somehow, and then use recursion to solve the smaller problem.
However, the main difficulty of doing recursion in this way is that we require the input graph to be a planar graph with a straight-line drawing. 
Thus we need to overcome two main obstacles to doing so: We need our base case to return a planar straight-line drawing, and we need a mechanism to glue together planar straight-line drawings.
We develop new sequential graph drawing algorithms to handle the base case, and new distributed graph drawing techniques to glue together smaller problems.

%\david{This is a very techinical note that discusses shaving a log factor off the total memory, I'm not sure this is particularly relevant}
% 
There are more technical issues that arise when applying the framework of Holm and T\v{e}tek~\cite{HolmT23}: Their framework partitions edges based on the location of one of the endpoints of the edge, even if the edge actually crosses many polygonal regions in the decomposition. Every region crossed by the edge needs to be aware of the existence of this edge when we recurse in order to guarantee that the drawing that is output has no crossings. If we directly use the partitioning construction of Holm and T\v{e}tek, the total sum of the sizes of the subproblems, including these edges that cross multiple regions, would increase by a factor of $O(\log n)$. We show how to avoid this extra factor in space with an improved divide-and-conquer scheme.

We also need to develop graph drawing algorithms to return planar straight-line embeddings of small pieces of graphs in polygonal regions found by our recursion. 
This itself is very challenging as these polygonal regions can have polygonal holes and degeneracies, and a straight-line embedding may not be possible.
We can instead relax this constraint to a drawing with $O(1)$ bends per edge, and we allow ourselves to move vertices around.
This is still not easy, as some vertices of the planar graph needed to be drawn are constrained to be placed on certain points in the boundary of the polygonal region. 
If the polygonal region has $O(s)$ sides, then we may need $O(s)$ bends to draw each edge of the planar graph.
This turns out to be far too large for our purposes. 
To overcome this issue, we will instead allow the algorithm to modify the polygonal region into one with $O(1)$ sides. 
However, this adds difficulties in the recursive step, where we now have to glue these modified polygons together.
We overcome these difficulties by using what we call the \emph{scaffold graph} which provides a way to glue our subproblems together.
%We believe that the novel techniques we develop here may also be of interest to the graph drawing community.

\paragraph{Overview of the algorithm.} 
For a graph $G$ with $n$ vertices and $m$ edges, we use a $(1/m^{\delta})$-cutting algorithm to partition the plane into $O(m^\delta)$ polygonal regions where each polygon has $O(1)$ sides and each polygon intersects at most $O(n^{1-\delta})$ edges of the input graph.
For each intersection between an edge of the input graph and an edge in the cutting, we add a virtual degree-2 vertex at the intersection point to subdivide the edge. We will bound the number of such subdivisions we perform. 
Given the cutting $\Xi$, let $G_{\Xi}$ be defined by the vertices and edges of the polygonal regions of $\Xi$. 
As  $|E(G_{\Xi})| = O(m^\delta)$, we can fit $G_{\Xi}$ into the local memory of a machine.

An $r$-division of a graph $G$ is a collection of $O(n/r)$ \emph{pieces}, where each piece is a subgraph of $G$ with at most $r$ vertices and $O(\sqrt{r})$ boundary vertices. 
We find an $n^{2 \delta / 3}$-division of $G_\Xi$.
We call this a \emph{cutting-division}, %an uninspired name 
as this is a partition by a division of a cutting. 
Each piece of the cutting-division corresponds to a subgraph of the old graph $G$, after the subdivision discussed above, with $O(n^{1-{ \delta / 3}})$ vertices and $O(n^{1-{2 \delta / 3}})$ boundary vertices. As there are  $n^{\delta / 3}$  pieces, the total number of boundary vertices is $O(n^{1-{ \delta / 3}})$. 
We note that the idea of taking planar separators of cuttings is not new, it was used in many papers before, e.g.~\cite{EppsteinGS10, ChazelleS11, Har-Peled14, HolmT23}.

We recurse on each piece, and the recursive call returns a solution for the piece and compresses the piece into a graph whose size is linear in the number of boundary vertices of the piece. We combine all the compressed graphs into a graph $\tG$. Due to the above bound on the number of boundary vertices, the size of $\tG$ is $O(n^{1-{ \delta / 3}})$. We also recurse on this graph. For this recursion to be possible, we need a straight-line drawing of $\tG$. For the base case, drawing a graph in a $h$-holed polygon with boundary complexity $c$ may require edges to be drawn with $O(c^h)$ bends per edge. Intuitively, this is due to the fact that the given graph topology may require some edges to be drawn with complicated homotopies that wind around all the holes. To make the drawing straight-line, each bend will be replaced with a degree-2 vertex. The complexity $O(c^h)$ is unacceptable if either $c$ or $h$ is non-constant.
To ensure that $h$ is constant, we will use a version of $r$-division~\cite{KleinMS13} which ensures that each region has $O(1)$ \emph{topological holes}. To ensure that $c$ is constant, we require that each of the $h$ holes forms a triangle. We devise a new algorithm for drawing the graph so the inside holes lie on a specific predetermined triangular boundary.
This involves building a \emph{scaffold} graph that we can draw as a planar straight-line drawing, and fix it so that boundary vertices lie on triangular boundaries of the polygonal region. This fixing is done by routing paths through an intermediate \emph{portal} vertex that is an intermediary between the triangular boundary, and the points that lie on the boundary.
%Throughout the recursion, it is crucial that we draw our graphs in polygons with $O(1)$ boundary complexity. To do so, we will consider a similar scaffolding approach when we combine the drawings returned by the recursive calls into a single drawing. Our method of fixing the drawing with portal vertices extends to the case where we have multiple pieces we need to join together all at once, and can be implemented in parallel.

\paragraph{Extensions of the framework.}
To extend our framework to give $(1+\eps)$-approximation for SSSP and shortest cycles, we use the $\eps$-emulator of Chang, Krauthgamer, and Tan~\cite{ChangKT22}. Given a planar graph $G$ with a set of terminals $T$, an \emph{$\eps$-emulator} is a planar graph $G'$ that contains all terminals in such a way that the distances between terminals in $G'$ are $(1+\eps)$-approximation of their distances in $G$. It was shown in~\cite{ChangKT22} that such an $\eps$-emulator of size $|V(G')| = \OO(|T|/\eps^{O(1)})$ exists and can be computed efficiently.
We show that we can compute $\eps$-emulators in $O(1)$ rounds in the MPC model, even though we cannot store all the terminals on one machine. We do so by applying our recursive framework.

To solve the shortest cycle problem, we categorize the cycles into ones that are confined to a single piece and ones that cross multiple pieces. The first type of cycle can be handled solely by recursion to each piece.
The second type of cycle can be handled by recursion to the graph resulting from combining what we call \emph{inside sparsifiers}, a compressed graph whose size is nearly linear in the number of boundary vertices such that the distances between the boundary vertices in the compressed graph are good approximations of their distances within the piece. Observe that $\eps$-emulators are exactly what we need for inside sparsifiers.

To solve SSSP, we will consider the following modified recursive framework where we recurse on each piece \emph{twice}. 
First, we recurse to find inside sparsifiers for each piece by constructing $\eps$-emulators in parallel.
Now we will solve the problem for the vertices of each piece in parallel. For each piece, we will build an \emph{outside sparsifier} on the vertices lying on the boundary of the piece from all the inside sparsifiers we already computed. 
We join the outside sparsifier to the old unsparsified graph within each piece and recurse to compute the shortest path from $s$.

It is well-known that the global minimum cut problem in a plane graph can be reduced to the shortest cycle problem in its dual graph. A similar reduction exists for the $st$-max-flow problem, which is equivalent to finding the shortest cycle separating two faces $s^*$ and $t^*$ in the dual.
While the dual graph construction is straightforward in the centralized setting, it is nontrivial in the MPC model as a face can have a complicated structure and as many as $O(n)$ sides.
The main difficulty in implementing this reduction in our setting is that in order to apply the tool sets that we have, it is required that the input graph is given a straight-line drawing. To circumvent the difficulty of constructing a straight-line drawing of the old dual graph, we consider a relaxed version of the problem where we allow each vertex in the dual graph to be represented by a connected subgraph, and we show that in this case, an $O(1)$-round construction algorithm in the MPC model is possible. This relaxed version is sufficient for the reduction discussed above. Our dual graph construction allows us to obtain an $O(1)$-round algorithm for the global minimum cut problem when combined with our shortest cycle algorithm.

The  $st$-max-flow problem is more challenging, as we need to maintain the faces of $s^*$ and $t^*$ in the dual. Instead of maintaining the entire face, which can be of arbitrary complexity, we will store a point in the faces corresponding to $s^*$ and $t^*$. Then, we use a similar type of recursion as we did for SSSP. %We will compute inside sparsifiers, and use that to compute certain outside sparsifiers. However, we will also 
When computing sparsifiers, we avoid sparsifying the piece that contains $s^*$ and $t^*$ to avoid changing the topology of paths relative to $s^*$ and $t^*$.

\subsection{Additional related work}\label{sec:related}

The 1-vs-2-cycles conjecture plays a central role in the computational complexity theory for the MPC model. Nanongkai and Scquizzato~\cite{nanongkai2022equivalence} demonstrated a large class of problems that are all equivalent to the 1-vs-2-cycles conjecture and studied the relation of this conjecture and other hardness assumptions. For example, 1-vs-2-cycles, planarity testing, minimum cut, bipartiteness, and counting connected components are equivalent in the sense that if any one of these problems can be solved in  $O(1)$ rounds with $\cS = n^{1-\Omega(1)}$, then all of these problems can be solved in $O(1)$ rounds with $\cS = n^{1-\Omega(1)}$.

Ghaffari, Kuhn, and Uitto~\cite{ghaffari2019conditional} showed that assuming the 1-vs-2-cycles conjecture, for any graph problem, any $\Omega(T)$ lower bound for the round complexity in the $\LOCAL$ model implies an $\Omega(\log T)$ lower bound for the round complexity of any algorithms in the MPC model with $\cS  =n^{1-\Omega(1)}$ that are \emph{component-stable}, which means that the output in a connected component is independent of all other connected components. This transformation of $\LOCAL$ lower bounds into MPC lower bounds has been successful in showing the conditional optimality of many existing MPC algorithms~\cite{balliu2023optimal,chang2019complexity,czumaj2021graph,ghaffari2020improved}.
The component-stability assumption needed in the transformation was shown to be \emph{necessary} by Czumaj, Davies, and  Parter~\cite{czumaj2021component}.

% Other distributed and parallel algorithms for planar graphs.
To the best of our knowledge,~\cite{HolmT23} is the only prior work that focuses on planar graphs in the MPC model. For the related $\LOCAL$ and the $\CONGEST$ models of distributed computing, there has been a large body of research designing distributed algorithms in planar networks: depth-first search~\cite{ghaffari2017near}, planarity testing~\cite{ghaffari2016planar,levi2021property}, separators~\cite{ghaffari2017near}, diameter~\cite{li2019planar}, reachability~\cite{parter2020distributed}, low-congestion shortcuts~\cite{GhaffariH16b}, and dominating set~\cite{alipour2020local,bonamy2021tight,lenzen2013distributed}.

% Connectivity/MST in general graphs in MPC/Congested clique?

%Graph connectivity is very well-studied in the MPC model. There are multiple algorithms~\cite{} that compute the connected components in $O(\log n)$ rounds in MPC. This bound has been  improved to $O(\log D) + O(\log_{m/n} \log n)$, nearly matching the lower bound $\Omega(\log D)$ assuming the 1-vs-2-cycles conjecture~\cite{Andoni2018,behnezhad2019near}. A very recent work~\cite{balliu2023optimal} showed that the term $O(\log_{m/n} \log n)$ can be removed when the underlying graph is a forest. 

\subsection{Roadmap}\label{sec:roadmap}

Here we describe the organization of the rest of the paper.
In \Cref{sec:prelim} we give the preliminaries.
In \Cref{sec:mpc_seperators} we describe our construction of \emph{cutting-divisions} in the MPC model that function as separators for embedded planar graphs.
After that, we present the necessary graph drawing operations we need in \Cref{sec:graph_drawing}, first by presenting an algorithm to draw the graph in the base case in a polygonal region of low complexity, and then a method of gluing graphs drawn in polygons of low complexity together.
In \Cref{sec:algorithms} we use our cutting-divisions and our graph drawing tools to build algorithms for embedded planar graphs.
In \Cref{sec:edit_distance}, we demonstrate an application of our results to edit distance computation.
To conclude, we end with some open questions in \Cref{sec:open_questions}.

\section{Preliminaries}\label{sec:prelim}

\paragraph{Massively parallel computing.}

In the MPC model, the input $m$-edge $n$-vertex graph is initially partitioned into machines with a memory of $\Theta(\cS)$ words of $O(\log n)$ bits. We will focus on the case where $\cS = n^\delta$ for any constant $\delta \in (0,1)$. 
The total number of machines is $\cM = O\left( \frac{m}{\cS}\right)$. We assume that these machines have unique identifiers in $[\cM]$. These machines communicate with each other in synchronous rounds.
For any routing task where each machine is the source and the destination of at most $O(\cS)$ messages of $O(\log n)$ bits, all the messages can be delivered to their destinations in one round of communication.

We can build a rooted tree $T$ over the $\cM$ machines so that each vertex of the tree has at most $\cS$ children and the tree height is $O(1/\delta) = O(1)$. Using this tree $T$, many tasks that require $\Omega(\log n)$ rounds to solve in CRCW PRAM can be solved in $O(1)$ rounds in the MPC model. For example, any machine can \emph{broadcast} a message of $O(\cS)$ words to all other machines in $O(1)$ rounds by first sending the messages to the root of $T$ and then broadcasting messages along the tree edges. To do so, we explain how a machine $u$ can send  $\cS$ words $w_1, w_2, \ldots, w_{\cS}$ to its $\cS$ children $v_1, v_2, \ldots, v_{\cS}$ in two rounds, as follows. In the first round, for each $i \in [\cS]$, $u$ sends $w_i$ to $v_i$. In the second round, for each $i \in [\cS]$ and each $j \in [\cS]$, $v_i$ sends $w_i$ to $v_j$. By the end of the second round, all machines $v_1, v_2, \ldots, v_{\cS}$ have received all words $w_1, w_2, \ldots, w_{\cS}$.

Moreover, given that each machine holds $O(\cS)$ items, it is possible to \emph{sort} all these items in $O(1)$ rounds using  $T$~\cite{goodrich2011sorting} in such a way that the first machine holds the  $O(\cS)$ smallest items, the second machine holds the $O(\cS)$ smallest items among the remaining items, and so on.

\paragraph{Planar graphs, subdivisions, and vertex splits.}
In this paper, we will only consider undirected planar graphs $G$.
We will let $V(G)$ denote the vertices of $G$ and $E(G)$ denote the edges of $G$.
Often we will consider weighted graphs, meaning that for every edge $e\in E(G)$, it has weight $w(e)$. If $e= (u,v)$, we will also write the weight as $w(u,v)$.
We will often also refer to this as the \emph{length} of an edge.

We will let $n = |V(G)|$ and $m = |E(G)|$.
The planar graphs we consider may not be simple. They may have multi-edges between two vertices.
We will assume that the graphs we consider have no isolated vertices, so by a simple counting argument $n\le 2m$. For simplicity, we will assume $n=\Theta(m)$ as well.
For all of the problems we will consider (e.g., connected components, minimum spanning forest, and shortest paths) isolated vertices are trivial to detect and handle separately. 

A \emph{subdivision} of an edge $e\in E(G)$ where $e=(u,u')$ for $u, u'\in V(E)$ is the deletion of $e$, followed by an insertion of a vertex $v_e$ with the two edges $e_1 = (u, v_e)$ and $e_2 = (v_e, u')$.
When we perform an edge subdivision, we will treat one of the edges, say $e_1$, as a canonical edge. This means that $w(e_1) = w(e)$, and we will often set $w(e_2) = 0$.
This way, the length of paths taking $e_1$ and $e_2$ is equal to the lengths of paths taking $e$.
This also preserves the value of the minimum spanning forest if all weights are positive. Furthermore, an edge is in the minimum spanning forest of the old graph if and only if the canonical edge is in the minimum spanning forest of the subdivided graph.

We will also consider performing \emph{vertex splits} of a vertex $v\in V(G)$.
If $v$ has incident edges $(v, u_1)$, $(v, u_2)$, $\ldots$, $(v, u_k) \in E(G)$, we partition the vertices $u_i$ into two sets $U_1$ and $U_2$. 
Then we delete $v$ and all its edges, and replace with two vertices $v_1$ and $v_2$, and add the edges $\{(v_1, u) \mid u\in U_1\}$ and $\{(v_2, u) \mid u\in U_2\}$ and the edge $(v_1, v_2)$.
We will treat one of the vertices, say $v_1$, as the canonical vertex.
Furthermore, we set $w(v_1, v_2) = 0$ and $w(v_1, u_1) = w(v, u_1)$ for $u_1\in U_1$ and $w(v_2, u_2) = w(v, u_2)$ for $u_2\in U_2$.
This way, the length of paths with $v$ as an intermediate vertex is preserved, as well as the value of the minimum spanning forest.

\paragraph{Polygonal regions and polygonal embeddings.}
A \emph{polygonal region with holes} $P$ or \emph{polygonal region} for short, is a connected planar polygon with one exterior boundary and zero or more interior boundaries.
We will call every connected region that is not the interior of $P$ a \emph{hole}. 
In particular, we call the connected region bounded by the external boundary of $P$ the \emph{exterior hole}, and the connected region bounded by one interior boundary an \emph{internal hole}.
We say $P$ is an \emph{$h$-holed polygonal region} if $P$ has $1$ exterior hole and $h-1$ interior holes.
The \emph{boundary} of $P$, denoted by $\partial(P)$, is the union of the boundary of each hole of $P$.
The \emph{boundary complexity} of $P$, denoted by $|\partial(P)|$, is the sum of the number of edges on the holes of $P$.
If every hole of $P$ has exactly $3$ sides, we say $P$ is a \emph{triangular polygonal region}.

\begin{figure}
    \centering
    \includegraphics[width=0.3\textwidth, page=3]{planar_separator.pdf}
    \hspace{3cm}
    \includegraphics[width=0.3\textwidth, page=1]{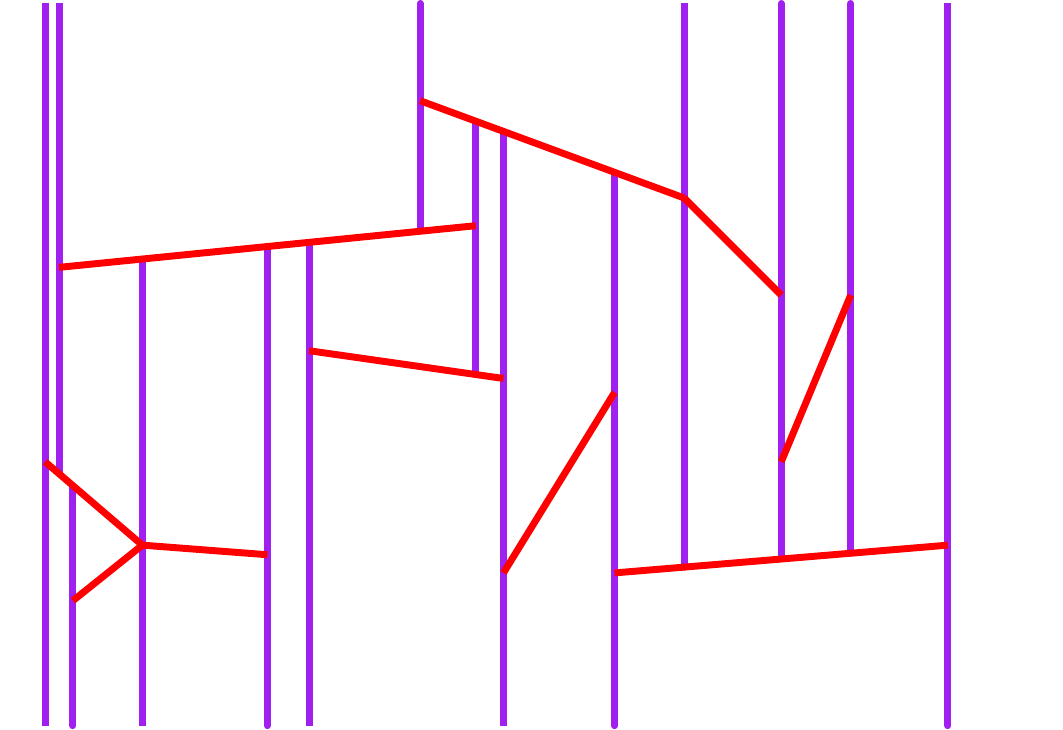}
    \caption{\textbf{(Left)} An $r$-division of a planar straight-line drawing inducing polygonal regions.
    Polygonal regions are allowed to be degenerate, like the blue region depicted, but they are guaranteed to be connected. 
    The purple piece has one yellow topological hole, which is also a polygonal hole.
    \textbf{(Right)} A vertical decomposition $\VD(S)$ of a set $S$ of non-crossing line segments. The line segments in $S$ are drawn in red.}
    \label{fig:sep_cutting}
\end{figure}

We allow the polygons to be degenerate. In particular, a single boundary can touch itself, and boundaries between holes can touch, as long as the polygon remains connected. This requirement stems from our use of planar separators. See \Cref{fig:sep_cutting} for an illustration.

%\paragraph{Polygonal embeddings}
Given a plane graph $G$ drawn in a polygon $P$ with $h$ holes, and some vertices $\nabla(G)\subseteq V(G)$ that lie on the boundary of $P$ that we refer to as \emph{terminals}, we say the triple $(G, \nabla(G), P)$ is a \emph{$h$-holed polygonal embedding} (or simply polygonal embedding). All graphs considered in this paper will be drawn with straight-line edges unless otherwise specified. We will sometimes deal with edges with bends. To handle these, we will add an additional vertex at every bend by subdividing the edge. 

\paragraph{Planar separators.}
Let $G$ be an $n$-vertex planar graph. 
For any integer $r\le n$, an \emph{$r$-division with few holes} \cite{Frederickson87, KleinMS13} 
(or simply, \emph{$r$-divisions}) is a collection of \emph{pieces} that are connected subgraphs of $G$ meeting the following conditions:
\begin{itemize}
    \item Each edge of $E(G)$ is in at least one piece.
    \item There are $O(n/r)$ pieces in the $r$-division.
    \item Each piece has at most $r$ vertices and $O(\sqrt{r})$ \emph{boundary vertices} (i.e., vertices that are also in some other piece).
    \item Any edge that is in two pieces is on a \emph{topological hole} (i.e., a face of a piece that is not a face of $G$) of each of them.
    \item Every piece has $O(1)$ topological holes.
\end{itemize}

This is the standard $r$-division of Klein, Mozes, and Sommer~\cite{KleinMS13}. They showed that it is possible to compute $r$-divisions in linear time for any $r$.
\begin{theorem}[$r$-divisions, Klein--Mozes--Sommer \cite{KleinMS13}] \label{thm:separators}
    Let $G$ be a biconnected embedded planar graph, and $r = \Omega(1)$. There is a linear time algorithm that computes an $r$-division of $G$ with few holes.
\end{theorem}

In the definition of $r$-divisions, topological holes are topological features of the pieces in relation to the old planar graph. 
We illustrate their connection to the geometric notion of holes for polygonal regions.
Observe that if $G$ is an embedded planar graph with each edge drawn as a straight line, then the union of the faces of a piece $R$ of a $r$-division that are faces of $G$ (i.e., not topological holes) forms a polygonal region with holes that we denote by $P_{R}$. The holes of $P_{R}$ correspond exactly to the topological holes of $R$, and $|\partial(P_R)| = O(\sqrt{r})$.
Note that these holes may not necessarily be simple, and may in fact be degenerate which we allow in our definition of polygonal regions.
See \Cref{fig:sep_cutting} for an illustration of this connection.

\paragraph{Vertical decomposition and cuttings.}\label{sec:vertical_decomposition}

We say a pair of line segments \emph{cross} if they intersect at a point that is not an endpoint of both line segments. Otherwise, we say that they are \emph{non-crossing}.
Let $S$ be a set of line segments. The \emph{vertical decomposition of $S$} (also known as the trapezoidal decomposition of $S$), denoted by $\VD(S)$, is constructed by adding the maximal possible vertical line segment that passes through the endpoint of each line segment and the intersection of a pair of line segments but does not intersect any of the other line segments. 
Observe that the $\VD(S)$ partitions the plane into (possibly unbounded and degenerate) \emph{vertical trapezoids}. 
See \Cref{fig:sep_cutting} for an illustration of $\VD(S)$ when $S$ is a non-crossing set of line segments.

\begin{definition}[$(1/r)$-cuttings]
    For a set $S$ of $n$ non-crossing line segments,
    a \emph{$(1/r)$-cutting} of $S$, denoted by $\Xi(S)$, is a partition of $\R^2$ into (possibly infinite) polygons $P_1, ..., P_k \in \Xi(S)$ with a constant number of sides such that each $P_i$ intersects at most $n/r$ line segments of $S$. The size of a $(1/r)$-cutting is $k$, the number of polygons.
\end{definition}

The next theorem about cuttings follows immediately from the $\eps$-net theorem of Haussler and Welzl \cite{HausslerW87}.
Holm and T\v{e}tek~\cite{HolmT23} showed that one can implement this in the MPC model.
\begin{theorem}[Haussler--Welzl \cite{HausslerW87}] \label{thm:cuttings_weak}
Let $S$ be a set of $n$ non-vertical non-crossing line segments and $0< \delta < 1$. Let $R \subseteq S$ be a random sample of size $m = O(r (\log r + \log \delta^{-1}))$. Then with probability at least $1-\delta$,  $\VD(R)$ gives a $(1/r)$-cutting of $S$. 
\end{theorem}

Optimal-sized cuttings were first presented by de Berg and Schwarzkopf \cite{BergS95}.
They considered the case where the line segments may cross.
The general idea is to take a vertical decomposition of a random sample of the desired cutting size, then refine the vertical trapezoids that have too many line segments crossing with a second round of sampling. 
We show in \Cref{sec:sublinear_cuttings_proof} an implementation of this result in the MPC model.
\begin{theorem}[Cuttings for line segments, de Berg--Schwarzkopf \cite{BergS95}] \label{thm:cuttings}
Let $H$ be a set of $n$ line segments that intersect at $\alpha$ points. There is a randomized algorithm to construct a $(1/r)$-cutting of $S$ of size $O(r + \alpha r^2/n^2)$ in time $O(n\log r + \alpha r/n)$. 
The algorithm can be made deterministic at the expense of increasing the runtime to $\poly(n)$.
\end{theorem}

For a set $S$ of $n$ non-crossing line segments, the number of intersection points is at most $n$, so \cref{thm:cuttings} gives an $O(n\log r)$ time algorithm to construct a $(1/r)$-cutting of size $O(r)$. Note that by definition any $(1/r)$-cutting must have size at least $\Omega(r)$, so these cuttings are optimal in size up to constant factors. 

\paragraph{Graph drawing.}

The problem of finding a straight-line embedding for planar graphs is well-studied. 
The first algorithm was presented by Tutte \cite{tutte1963draw} in 1963.
Given a planar graph $G$ with $n = |V(G)|$ vertices, we will focus on drawings that place vertices on an $O(n)\times O(n)$ grid as these will keep the bit complexity of our embeddings low.
Schnyder \cite{schnyder1989planar, Schnyder90} was the first to discover a construction of a straight-line drawing of any planar graph in such a grid.

\begin{theorem}[Schnyder \cite{Schnyder90}] \label{thm:graph_drawing}
Any plane graph $G$ with $n\ge 3$ vertices has a straight-line embedding in an $(n-2)\times (n-2)$ grid.
Furthermore, we can compute the drawing in $O(n)$ time.
\end{theorem}

\paragraph{\texorpdfstring{$\eps$}{Eps}-emulators.}

For a weighted undirected connected planar graph $G$ with a special set of vertices $T$ called \emph{terminals}, we refer to $(G, T)$ as an instance. We call an instance a \emph{planar instance} if the graph $G$ is planar. A planar instance is a \emph{$h$-hole instance} for an integer $h > 0$ if the terminals lie on at most $h$ faces in the embedding on $G$. We will also refer to these faces as \emph{topological holes}.
Given an instance $(G, T)$, an \emph{$\eps$-emulator} for $(G, T)$ is a planar instance $(G', T)$ such that for all $x, y \in T$:
\[ \dist_G(x, y) \le \dist_{G'}(x, y) \le (1+\eps)\cdot \dist_G(x,y). \]

Chang, Krauthgamer, and Tan \cite{ChangKT22} proved that $\eps$-emulators can be computed in near-linear time $\OO(n)$\footnote{In this paper we use $\OO(\cdot)$ to suppress polylogarithmic factors.} for constant $\eps > 0$.
\begin{theorem}[Theorems 1.1 and 5.8, Chang--Krauthgamer--Tan~\cite{ChangKT22}] \label{thm:eps_emulator}
    For every instance $(G, T)$ and a parameter $0<\eps < 1$ with $n = |V(G)|$ and $k=|T|$, 
    an $\eps$-emulator $(G', T)$ of size $|V(G')| = \OO(k/\eps^{O(1)})$ can be computed in $\OO(n/\eps^{O(1)})$ time.
    Furthermore, if $(G, T)$ is an $h$-hole instance, $(G', T)$ is also an $h$-hole instance, and $G$ and $G'$ have the same order of vertices around each topological hole.
    %If $k \le n/\log^D n$ for a large enough constant $D$, and the range of weights are bounded by a polynomial in $n$, then this can be done in $O_\eps(n)$ time. 
\end{theorem}

\section{MPC algorithm for separators in embedded planar graphs} \label{sec:mpc_seperators}

In this section, we show how to construct a version of a separator for embedded planar graphs. These separators will be constructed from graph divisions on geometric cuttings that we call \emph{cutting-divisions}. We begin by presenting a sublinear-time algorithm for constructing a cutting of line segments in \Cref{sec:sublinear}. Then we show how this randomized sublinear-time algorithm implies a method of computing cuttings from a random sample in \Cref{sec:cutting-divisions}. Afterward, we describe how we put these cuttings together with graph divisions to construct a cutting-division that functions as a separator in \Cref{sec:mpc_cutting-division}.

\subsection{Sublinear and MPC algorithms for constructing cuttings} \label{sec:sublinear}
It is possible to compute $(1/r)$-cuttings for a set $S$ of $n$ non-crossing line segments in sub-linear time.
The details are technical but standard, so we defer the proof to \Cref{sec:sublinear_cuttings_proof}.

\begin{restatable}[Sublinear time $(1/r)$-cuttings of size $O(r)$]{theorem}{sublinearcuttings}
\label{thm:sublinear_cuttings}
Let $S$ be a set of $n$ non-crossing line segments and $0< \delta < 1$. 
Let $R \subseteq S$ be a random sample of size $m = O(r(\log r + \log \delta^{-1}))$. 
Then there exists a randomized algorithm that takes as input the sample $R$, constructs a $(1/r)$-cutting of $S$ of size $O(r)$ in time $O(m\log r)$ and succeeds with probability at least $1-\delta$. In particular, a $(1/r)$-cutting of $S$ of size $O(r)$ can be constructed from a sample of size $O(r\log n)$ in $O(r\log n \log r)$ time with high probability.
\end{restatable}

\paragraph{Remark.}
This result on cuttings applies even if the line segments cross.  
The proof of \Cref{thm:sublinear_cuttings} easily extends to any family of line segments (or even general curves) with a dependency on the number of crossings given by \Cref{thm:cuttings}.
These results also extend beyond line segments to reasonably ``nice'' families of objects with the number of crossings between every pair of objects bounded by a constant,
as the cuttings of \Cref{thm:cuttings} can be adapted to any such objects with bounded VC-dimension.
These techniques easily extend to the differentiable curves with bounded total curvature considered by Holm and T\v{e}tek \cite{HolmT23}. 
We opt to abstain from presenting precise bounds, as the technical intricacies and particular proofs are contingent upon the specific types of objects under consideration.

\subsection{Cutting-divisions}\label{sec:cutting-divisions}
\begin{figure}
    \centering
    \includegraphics[width = 0.33\textwidth, page=3]{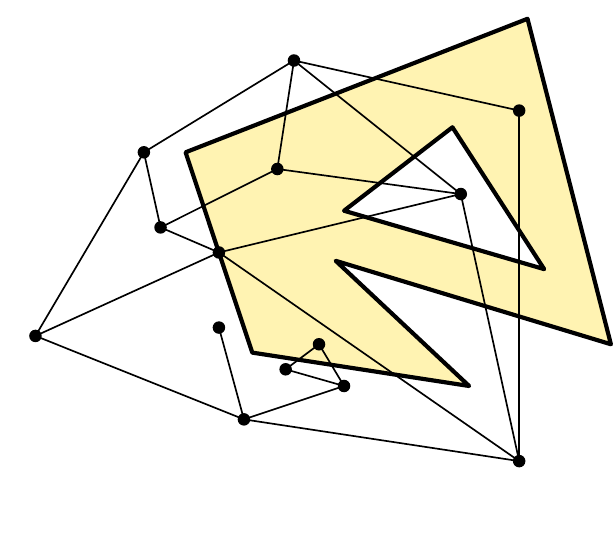}
    \hspace{-3em}
    \includegraphics[width = 0.33\textwidth, page=5]{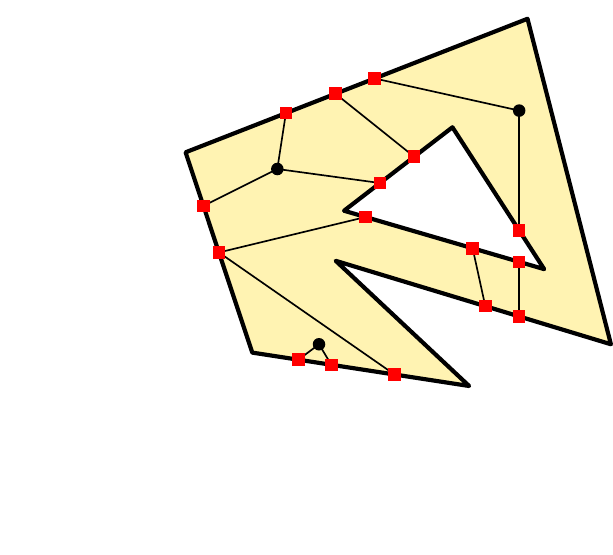}
    \hspace{3em}
    \includegraphics[width = 0.31\textwidth, page=7]{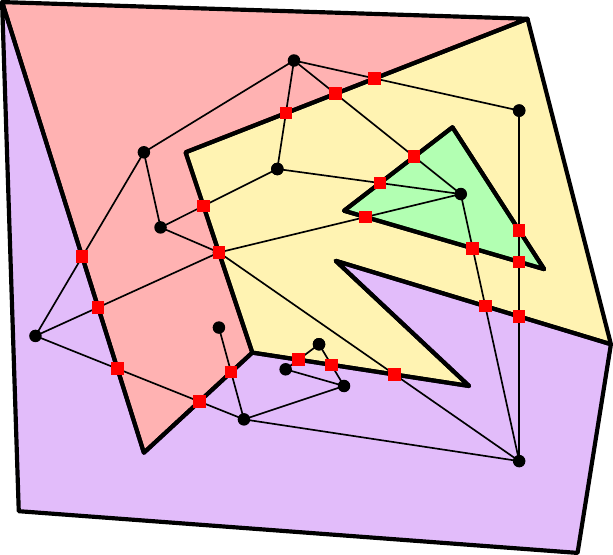}
    \caption{
    \textbf{(Left)} An embedded planar graph $G$ and a polygonal region $P$. 
    \textbf{(Middle)} $G_P$, the $P$-induced subgraph of $G$. The vertices marked by red squares form $\nabla(G_P)$. 
    \textbf{(Right)} A $\Pi$-induced subdivision of $G$ for a partition $\Pi$. The places where edges are subdivided are marked by red squares.}
    \label{fig:induced_subgraph}
\end{figure}

Let $G$ be an embedded planar graph, and $P$ be a polygonal region. We define the \emph{geometrically induced subgraph} of $G$ by $P$ (or \emph{$P$-induced subgraph}) to be the graph $G_P$ obtained by restricting $G$ to $P$. To be precise $G_P$ consists of all edges of $E(G)$ that lie completely in $P$, and for every edge $e\in E(G)$ that crosses the boundary of $P$, we add a vertex at every location where the edge crosses $P$, and add an edge between consecutive vertices along the edge in the interior of $P$. Note that the old endpoints of the edge in $G$ may not lie in $P$. Also, a single edge of $G$ may be divided into multiple pieces even if $G$ is a $1$-holed polygon.

Let $\Pi$ be a partition of the plane into polygonal regions $P_1, P_2, ..., P_k$. We define the \emph{partition induced subdivision} of $G$ by $\Pi$ (or \emph{$\Pi$-induced subdivision}) as the graph $G_{\Pi}$ obtained by taking the union of $G_{i}$ the $P_i$-induced subdivisions of $G$ for all $i=1,...,k$. See \Cref{fig:induced_subgraph} for an illustration of these definitions.

For a polygonal embedded graph $(G, \nabla(G), P)$, a partition $\Pi$ of $P$ induces polygonal embedded graphs $(G_i, \nabla(G_i), P_i)$ where $G_i$ is the $P_i$-induced subgraph of $G$.
Due to subdivisions, a geometrically induced subgraph $G_i$ might not be a subgraph of $G$ in general (see \Cref{fig:induced_subgraph}), but $G_i$ is a subgraph of $G_{\Pi}$.

In the subsequent discussion, we will define \emph{$(s, 1/r)$-cutting-divisions} for $O(1)$-holed polygonal embeddings $(G, \nabla(G), P)$ with $n$ vertices and $m$ edges.
The name is derived from the fact that they are $s$-divisions of a graph constructed from $(1/r)$-cuttings where $0< s\le r \le m$.
After that, we will describe the construction from a random sample of size $O(r\log n)$ in detail in \Cref{lem:cutting-division} and show how to do this in $O(1)$ rounds in the MPC model in \Cref{lem:mpc_cutting-division}.

In this paper we will typically have $r = \cS^\alpha$ for a suitably small constant $0 < \alpha <1$, $s = r^{2/3}$, and $S = n^{\delta}$ for some $\delta > 0$.
These parameters are chosen so that we can construct cutting-divisions in one machine.
An $(s, 1/r)$ cutting-division $\Gamma$ is a partition of $P$ into $O(1)$-holed polygonal regions $P_1,..., P_k$. 
There are $k = O(r/s)$ polygonal regions, each of which has a relatively small boundary. 
Each region balances the number of edges of $G$ that lie in the region to be $O(m/k)$, and has a much smaller number of edges that cross the boundary. 
The polygons themselves have a fairly small number of total sides so that they can be distributed to all machines. 

The following definition may look quite complicated, but is in fact a list of the combined properties of the division and the cutting.
\begin{definition}[$(s, 1/r)$-cuttings]
Let $s$ and $r$ be two parameters with $s\le r$. An $(s, 1/r)$-cutting-division $\Gamma$ of a $O(1)$-holed polygonal embedding $(G, \nabla(G), P)$ with $n$ vertices and $m$ edges is a partition of $P$ into disjoint polygonal regions $P_1, P_2, \ldots, P_k \in \Gamma$. This gives a $\Gamma$-induced subdivision $G_\Gamma$ and the polygonal embedded graphs $(G_i, \nabla(G_i), P_i)$ for $i=1,...,k$ with these properties:
\begin{itemize}
\item The number of polygonal regions is $k = O(r/s)$.
\item Each polygonal region $P_i\in \Gamma$ has $O(1)$ holes.
\item $|\partial(P_{i})| \le |\partial(P)| + O(\sqrt{s})$ for each region $P_i\in \Gamma$.
\item $\sum_{i=1}^k |\partial(P_i)| = |\partial(P)| + O(r/\sqrt{s})$.
\item $|E(G_i)| \le ms/r$.
\item $\sum_{i=1}^k |\nabla(G_i)| = |\nabla(G)| + O(m/\sqrt{s})$.
\item $|V(G_{\Gamma})| = n + O(m\sqrt{s}/r)$, i.e., we introduce at most $O(m\sqrt{s}/r)$ vertices from subdivisions.
\end{itemize}
\end{definition}

%\yijun{The paragraph below is a bit repetitive: the issue of $m$ vs $n$ has been discussed in the preliminary section. The $O(m)$ bound on the number of additional edges due to subdivisions is discussed twice.} \david{fixed?}

The above definition is very similar to the definition of \emph{pseudodivisions} used by Holm and T\v{e}tek \cite{HolmT23}.
We point out the differences in these definitions. 
The biggest difference is that we subdivide edges at every intersection of the edge with the partition $\Gamma$.
This is necessary for us, as we will recurse in each polygonal region and the recursive calls produce drawings. If there are edges that are not considered in the recursive case, then it is very difficult to avoid crossings when adding them back. 
Furthermore, the subdivision does not increase the number of edges $m$ in the graph by too much. To be precise, with the sublinear time cutting algorithm of \Cref{thm:sublinear_cuttings} producing optimally sized cuttings, the total number of edges due to subdivisions can be upper bounded by $O(m)$.
Another difference is that we do not take the dual when computing the division.  Instead, we directly apply divisions to the graph formed by the cutting and subdivide every edge for every boundary that it crosses.
Note that doing so will not significantly increase the number of edges in the graph, as long as 
we do not have too many boundaries. The number of boundaries we will have is relatively small, and the cutting guarantees that not too many edges are in each region, and thus not too many edges cross each boundary.
%
%One final difference is that we will not worry about the number of vertices. Indeed, if the graph has no isolated vertices, then $n = \Theta(m)$. 
%If we do have isolated vertices, then we can easily detect and handle them depending on the problem. For example, for counting connected components, we can remove all the isolated vertices after counting the  number of the isolated vertices.

We show that we can construct $(s, 1/r)$-cutting-divisions of polygonal embeddings.
\begin{lemma}[$(s, 1/r)$-cutting-division construction] \label{lem:cutting-division}
Let $(G, \nabla(G), P)$ be a polygonal embedding of a planar graph with $m$ edges and let $s$ and $r$ be parameters with $s\le r$ and $|\partial(P)| \le O(r)$.
There exists an algorithm, that given $P$ and a sample of $O(r\log m)$ edges from $G$, constructs $\Gamma$ an $(s, 1/r)$-cutting-division of $(G, \nabla(G), P)$.
\end{lemma}
\begin{figure}
    \centering
    \includegraphics[width=0.30\textwidth, page=3]{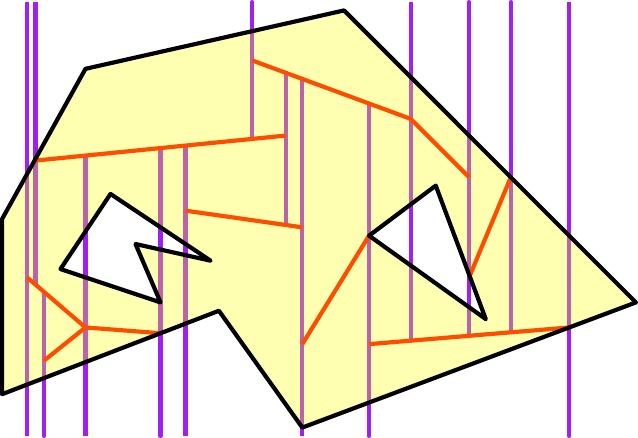}
    \includegraphics[width=0.30\textwidth, page=4]{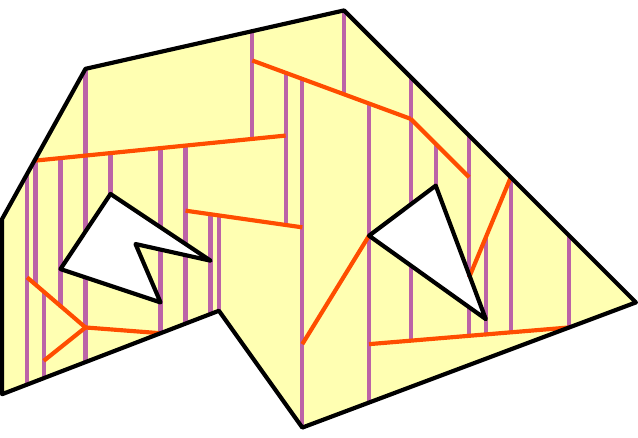}
    \includegraphics[width=0.30\textwidth, page=5]{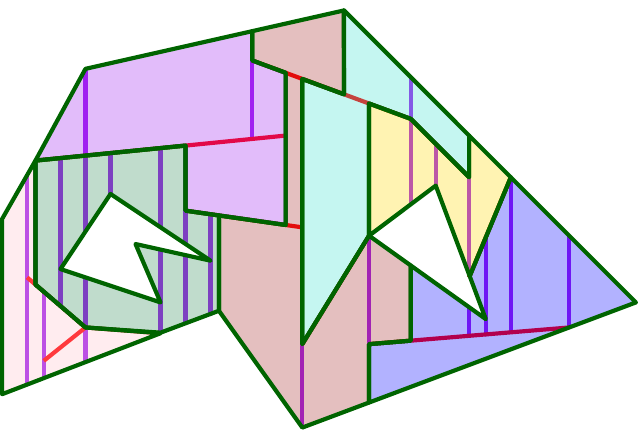}
    \caption{\textbf{(Left)} Construction of cutting-division  by constructing a cutting.
    \textbf{(Middle)} Clipping the cutting to the polygon and taking a vertical decomposition.
    \textbf{(Right)} An illustration of a division with few holes.}
    \label{fig:cutting-division}
\end{figure}
\begin{proof}
Begin by constructing a $(1/r)$-cutting of the edges of $G$ from the $O(r\log m)$ sample as well as the edges of $\partial(P)$ by using \Cref{thm:sublinear_cuttings} and let $\Xi_0$ denote the cutting that results from the theorem. Clip regions that extend outside of $P$ and take the vertical decomposition of the clipped regions. Let this resulting cutting be $\Xi$ which has size $O(r)$ since $\Xi_0$ and $\partial(P)$ have size $O(r)$.

Consider the cutting $\Xi$ as an embedded planar straight-line drawing of a graph $G_{\Xi}$ 
with vertices at the endpoint of every line segment of $\Xi$, and edges between vertices connected by line segments.
Note that $|E(G_{\Xi})| = O(r)$ since $\Xi$ had size $O(r)$. Using \Cref{thm:separators}, we can construct $\Gamma$ an $s$-division of $G_\Xi$.

We claim that $\Gamma$ is an $(s, 1/r)$-cutting-division as desired. 
First, we view $\Gamma$ as a partition of $P$, and use the fact that it was constructed from a $s$-division of $\Xi$. 
There are $k=O(r/s)$ polygonal regions $P_1, ..., P_k \in \Gamma$, each of which is constructed from pieces that have at most $s$ internal vertices and $O(\sqrt{s})$ boundary vertices, and $O(1)$ holes.
This implies that each piece has $O(s)$ trapezoids of the cutting $\Xi$ and $O(\sqrt{s})$ boundary edges by a simple application of Euler's formula to each piece.
Note that this does not count the boundaries of $P$ which may be a part of any of the $P_i$ regions.
This gives us the bound of $|\partial(P_i)| \le |\partial(P)| + O(\sqrt{s})$, and $\sum_{i=1}^k |\partial(P_i)| = |\partial(P)| + O(r/\sqrt{s})$.

Now we use the fact that the underlying graph $\Xi$ is a $(1/r)$-cutting of $P$.
Since each piece corresponding to $P_i$ has at most $O(s)$ trapezoids in the interior and $O(\sqrt{s})$ boundary edges, and each trapezoid or edge of the cutting intersects at most $O(m/r)$ edges of $G$, this means that $P_i$ contains or intersects at most $O(ms/r)$ edges of $G$. 

Now let us bound $\nabla(G_i)$ the number of vertices of $G_i$, the $P_i$-induced subgraph of $G$ that lies on the boundary of $P_i$. 
The boundary of $P_i$ consists of at most $O(r/\sqrt{s})$ edges of $G_\Xi$, each of which has at most $n/r$ edges of $G$ crossing, or was part of $\partial(P)$ which overall had $\nabla(G)$ vertices. 
This implies that the number of total boundary vertices, summed across all $G_i$ is $\sum_{i=1}^k |\nabla(G_i)| = |\nabla(G)| + O(m/\sqrt{s})$.
Furthermore, we subdivide edges crossing the edges of $G_\Xi$ used by $\Gamma$, so we subdivide $O(m/\sqrt{s})$ edges and increase the number of vertices and edges by that much.
Note that when we have edges $G$ on the boundary of $\Gamma$, we can arbitrarily assign it to the region that lies below (though the endpoints will still be boundary vertices in all adjacent regions).
\end{proof}

\paragraph{Remark.} 
We remark that it is possible to strengthen the definition of $(s, 1/r)$-cutting-divisions.
This can be done by modifying the $s$-division theorem of Klein--Mozes--Sommer~\cite{KleinMS13} that alternate between finding cycle separators that decrease the size of the boundary, decrease the number of holes, and decrease the number of vertices in each piece of the division. 
For instance, we could guarantee that $|\partial(P_i)| \le O(\sqrt{s})$, provided that $\partial(P) \le O(r/\sqrt{s})$ by also alternating between finding cycle separators that decrease the boundary of $\partial(P)$. We can also guarantee that $|\nabla(G_i)| \le O(m/\sqrt{s})$, provided that $\nabla(G) \le O(m/\sqrt{s})$ by also alternating between finding cycle separators that decreases the size of $\nabla(G)$. 
This is not needed for our algorithms but may be of independent interest and may be useful in other applications. 

\subsection{MPC algorithm for computing cutting-divisions}
\label{sec:mpc_cutting-division}
The following theorem is a consequence of plugging in our modified $r$-cutting construction in the algorithm of Holm and T\v{e}tek \cite{HolmT23} with some differences.
For completeness, we describe the entire algorithm.

\begin{lemma}[MPC algorithm for constructing cutting-divisions] \label{lem:mpc_cutting-division}
Let $(G, \nabla(G), P)$ be a polygonal embedding with $n$ vertices, $m$ edges. 
Let our parameters $r$ and $s$ satisfy $s\le r$, $|\partial(P)| \le O(r)$, and that a $(s, 1/r)$-cutting-division fits in $S$ memory,
then there exists an algorithm that takes $(G, \nabla(G), P)$ 
as input computes an $(s, 1/r)$-cutting-division $\Gamma$  
such that the layout of edges in memory satisfies that:
\begin{enumerate}[label=(\arabic*)]
    \item each edge that crosses a boundary is subdivided into contiguous pieces 
    \item each machine stores edges from one polygonal region, and the polygonal region itself
    \item each region is stored in consecutive machines
    \item the first machine stores the range of machines that each region is stored in
\end{enumerate}
The algorithm uses $O(\cS)$ space per machine, $O(n/\cS)$ machines, and performs $O(1)$ rounds in expectation and with high probability.
\end{lemma}
\begin{proof}
    Each machine samples its stored edges with probability $O(r\log m /m)$ and sends them to machine $1$.
    Machine $1$ can compute a $(s,1/r)$-cutting-division of $\Gamma$ by \Cref{lem:mpc_cutting-division}, and can broadcast the $\Gamma$ to all machines.

    Each machine needs to subdivide the edges that cross boundaries of $\Gamma$. 
    However, since each machine stores $O(\cS)$ edges, and each edge may cross $O(r)$ boundaries, 
    we cannot do the subdivision on one machine as that may result in as many as $O(r\cS)$ total edges.
    Instead, each machine will first count for each edge $e\in E$ how many regions that edge intersects $\ell_e$ with an algorithm we will describe in the following paragraph.
    Then, it will request for $t = O(\sum \ell_e / \cS)$ machines and perform a weighted partition of the edges among the $t$ machines so each machine can hold $O(\cS)$ edges after the subdivision. This can be done in $O(1)$ rounds.
    
    To detect how many times a collection of $O(\cS)$ line segment intersects the $O(r)$ boundaries of $\Gamma$ is exactly counting bichromatic intersections between red and blue line segments where there are no red/red or blue/blue intersections (except at endpoints). This can be done in $O(\cS \log \cS)$ time and $O(\cS)$ storage for $O(\cS)$ red and $O(\cS)$ blue line segments using the algorithm of 
    Theorem 3.1\footnote{Technically the theorem states that it is possible to count the \emph{total} number of intersections but can be easily adapted to count the number of intersections for each line segment.}
    of Chazelle--Edeslbrunner--Guibas--Sharir~\cite{ChazelleEGS94}. 

    Each machine can build the point location data structure 
    of Kirkpatrick~\cite{Kirkpatrick83}
    that can determine which region each edge lies in $O(\log r)$ time, using $O(r)$ total storage.
    Using this data structure, the machines can now sort the edges so that consecutive machines store edges for one region. This sort can be done in $O(1)$ rounds.
\end{proof}

We note that the above theorem can be made deterministic by derandomizing the cutting construction. This can also be done using standard techniques which we give a proof of in \Cref{sec:deterministic_cuttings}.
\begin{restatable}[Deterministic MPC Cuttings]{theorem}{thmdetcuttings} \label{thm:det_mpc_cuttings}
There exists a deterministic MPC algorithm for computing a $(1/r)$-cutting of $n$ line segments $S$ for $r=\cS^{\alpha}$ for a sufficiently small $\alpha >0$ that uses $O(1)$ rounds with $O(n/\cS)$ machines with $O(\cS)$ memory per machine where $\cS = n^\delta$ for some constant $\delta > 0$.
\end{restatable}

\section{Graph drawing} \label{sec:graph_drawing}

\begin{figure}
    \centering
    \includegraphics[width=0.99\textwidth, page=3]{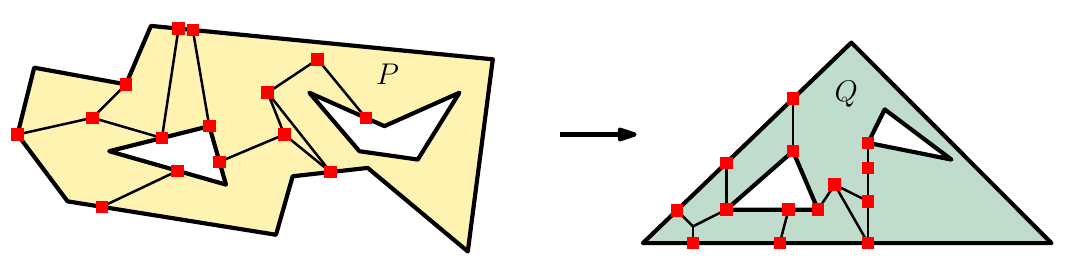}
    \includegraphics[width=0.99\textwidth, page=3]{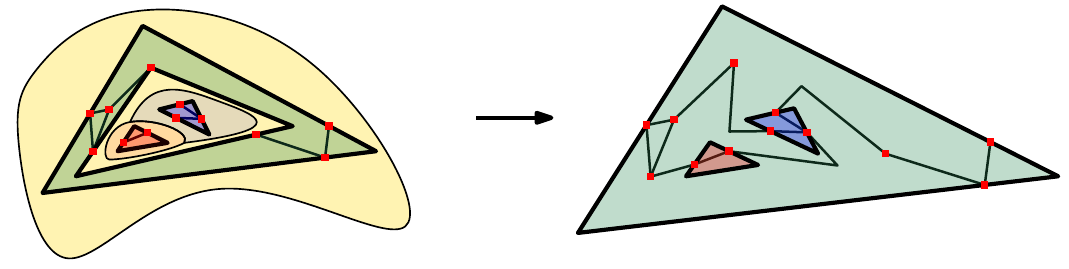}
    \caption{
    \textbf{(Top)} Objective 1: Drawing of a graph $(G, \nabla(G), P)$ into a polygonal region $Q$.
    \\
    \textbf{(Bottom)} Objective 2: Given a collection of graphs, one for each polygonal region, we need to glue the boundary vertices corresponding to the same vertex of $G_\Pi$ together with an edge (possibly with bends). }
    \label{fig:gd_objectives}
\end{figure}

In this section, we prove various graph drawing lemmas that we need to compute explicit graphs in order to use recursion in our MPC algorithms.
Throughout this section, we will assume that the input polygonal regions are non-degenerate. However, we cannot give this guarantee as the polygons resulting from our cutting-divisions may be degenerate. With more care, we can reduce the degenerate case to the non-degenerate case, which we will discuss in \Cref{sec:degenerate}.

We list our main two objectives of this section and illustrate them in \Cref{fig:gd_objectives}. The first objective is for the base case of the recursion of our MPC algorithms. The second objective is for handling the recursive case.
\begin{description}
    \item[1. Redrawing.] Draw a polygonal embedded graph $(G, \nabla(G), P)$ in a triangular polygonal region $Q$ with the same number of holes as $P$. (\Cref{lem:redrawing})
    \item[2. Gluing.] Given a partition $\Pi$ of a polygonal embedded graph $(G, \nabla(G), P)$ and the induced partition into $(G_i, \nabla(G_i), P_i)$ for $i=1,..., k$, as well as a collection of graphs $\cG = \{(G_i', \nabla(G_i), Q_i) \mid i=1, ..., k\}$ with the same boundary vertices but possibly different graphs drawn in different polygonal regions,
    \emph{glue} the boundary vertices of each graph of $\cG$ that correspond to copies of the same vertex in $G_{\Pi}$, i.e., connect them with edges. (\Cref{lem:gluing_lemma})
\end{description}

To accomplish the first objective, we show that if $P$ and $Q$ are triangular polygonal regions (a polygonal region where each boundary is a triangle) with $O(1)$ holes, then we can find a piecewise linear transform that allows us to obtain $(G, \nabla(G), Q)$ from $(G, \nabla(G), P)$ while adding a constant number of edges to each straight-line edge. This algorithm is easily made parallel as the same transform is applied to each edge independently.
For this result to be useful, we prove the Redrawing Lemma in \Cref{sec:graph_redrawing}, which shows that if $P$ is not a triangular polygonal region, then we can redraw a graph $(G, \nabla(G), P)$ as $(G,\nabla(G), Q)$ where $Q$ is a triangular polygonal region.
This algorithm is purely sequential and used in the base case of our MPC algorithms.
For the second objective, we need some mechanism of \emph{gluing} together pieces of graphs we obtain from applying recursion with $(s, 1/r)$-cutting-divisions. To accomplish this, we prove the Gluing Lemma in \Cref{lem:gluing_lemma} that says we can do so while adding $O(1)$ bends and subdividing $O(1)$ edges. Furthermore, we can do this in the MPC model in $O(1)$ rounds.

\paragraph{Challenges.} We illustrate the challenges faced in trying to solve the first objective.
There are known graph drawing algorithms like Tutte embeddings \cite{tutte1963draw} that allow us to pick any convex placement of the vertices of the outer face and draw the rest of the edges of the planar graph with non-crossing straight lines.
However, such graph drawing methods do not allow for any control for us to draw the interior holes.
In particular, there is no way to guarantee that the interior holes are on the edges of a triangle, especially if there are many vertices on that interior hole. 
To do this we need to allow for some number of bends in the edges.
We also need to output exact coordinates with bounded bit complexity that we will use later.

The second objective also comes with its own challenges.
We need to ensure that we compute the entire gluing all at once, since we need to implement this algorithm in the MPC model with $O(1)$ rounds.
However, the partition can have polygonal regions arbitrarily nested, boundary vertices of high degree, or polygonal regions adjacent to many other polygonal regions. 
Simultaneously, we are limited in the operation we are allowed to do since we cannot store the entire graph on one machine, and we also need to be careful with the bit complexity of the coordinates since the amount of total memory we have is bounded.

\paragraph{Our techniques.}
At a high level, we will handle both of these objectives with the same method:
\begin{center}
\textit{old polygonal embedding $\rightarrow$ Scaffold graph $\rightarrow$ New polygonal embedding}
\end{center}
The \emph{scaffold graph} is an abstract planar graph that will facilitate our transition from the old polygonal embedding to a new one.

We will describe this in more detail for the first objective.
For simplicity, we will slightly relax the goal of computing the drawing $(G, \nabla(G), Q)$,
and instead compute $(G', \nabla(G), Q)$ for a graph $G'$ that contains $G$ as a minor\footnote{This is not actually necessary: 
With more care, it is possible to transform a drawing of $G'$ into $G$ by adding constantly many more bends per edge.
However our MPC algorithms already involve subdividing edges and performing vertex splits, so this does not make a difference to us.}
(i.e., there exists a set of edges of $G'$ that can be contracted to get $G$).
We describe how we construct a \emph{scaffold graph} from this graph.
The idea is that we will insert three points arranged in a triangle $T$ on the inside of every hole $H$ of $P$. This will eventually form the sides of the triangular boundary of a triangular polygonal embedding $Q$.
However, we need to ensure that we have some mechanism of connecting points on the actual hole $H$ to points along $T$.
To do so, we add what we call \emph{portal} vertices between $H$ and $T$ and connect them to the vertices of $H$ and $T$ with edges. 
Then we prove the Boundary Routing Lemma in \Cref{lem:boundary_routing} to show that we can draw non-crossing paths with at most $3$ bends from anywhere on $H$ to the corresponding location on $T$. 
Thus this scaffold graph provides the necessary portal vertices to allow us to draw $G'$ and is described in \Cref{sec:boundary_routing}.

Surprisingly, these two lemmas are the only ingredients necessary to accomplish the second objective and allow us to prove the Gluing Lemma of \Cref{lem:gluing_lemma}.
In this case, we need to compute a scaffold graph for a cutting-division that we define in \Cref{sec:gluing}.
The main idea is for every polygonal region in the cutting division, we replace each of its boundaries with a triangle on the \emph{inside} of the region (along with the portal vertices). 
Using this scaffolding, if we had redrawings of each polygonal region into a triangular region, we can essentially \emph{glue} the graphs of each redrawing together. We explain this in detail in \Cref{sec:gluing}.

\subsection{Compatible triangulations and graph morphing} \label{sec:triangle_morph}

 Given a polygonal region $P$ with holes and a homeomorphic polygonal region $Q$ (a region with the same number of holes, and each boundary having the same number of sides), we say that the triangulations $\cT_P$ and $\cT_Q$ of $P$ and $Q$ respectively are \emph{compatible triangulation} if there is a bijection $f$ from the vertices of $\cT_P$ to the vertices of $\cT_Q$ such that $(p, p', p'')$ is a triangle of $\cT_P$ if and only if $(f(p), f(p'), f(p''))$ is a triangle of $\cT_Q$. 
 Babikov, Souvaine, and Wenger~\cite{BabikovSW97} showed the following result. 
 
\begin{theorem}[Babikov--Souvaine--Wenger~\cite{BabikovSW97}] \label{thm:compatible_triangulations}
Let $P$ and $Q$ be two homeomorphic labeled polygonal regions with holes with vertex sets $\{p_1, ..., p_n\}$ and $\{q_1, ..., q_n\}$ such that there is a homeomorphism from $P$ to $Q$ mapping $p_i$ to $q_i$. Compatible triangulations and piecewise-linear homeomorphisms of $P$ and $Q$ matching $p_i$ and $q_i$ can be constructed in $O(n^2)$ time using $O(n^2)$ triangles.
 \end{theorem}

This allows us to prove the following theorem.
\begin{lemma}[Triangular morphing lemma] \label{lem:morph}
Given an embedded planar graph $(G, \nabla(G), Q')$ where $Q'$ is a triangular polygonal region, and another triangular polygonal region $Q$ with the same number of holes, we can compute the embedded planar graph $(G, \nabla(G), Q)$ by applying a piecewise linear homeomorphism $\phi$ from $Q$ to $Q'$ to each edge. If $Q'$ had $O(h)$ holes, $\phi$ decomposes into $O(h^2)$ linear pieces and each edge of $G$ becomes an edge with at most $O(h^2)$ bends.
\end{lemma}
\begin{proof}
By \Cref{thm:compatible_triangulations}, we can compute a compatible triangulation between $Q'$ and $Q$ and a linear homeomorphism $\phi$ of size $O(h^2)$. When we apply this linear homeomorphism to $G'$, each straight-line edge $e\in E'$ drawn on $S'$, results in $\phi(e)$ drawn on $S$ as straight-line edges with at most $O(h^2)$ bends.
\end{proof}

\subsection{Scaffold graphs for a polygonal region}
\begin{figure}
    \centering
    \includegraphics[width=0.4\textwidth, page=2]{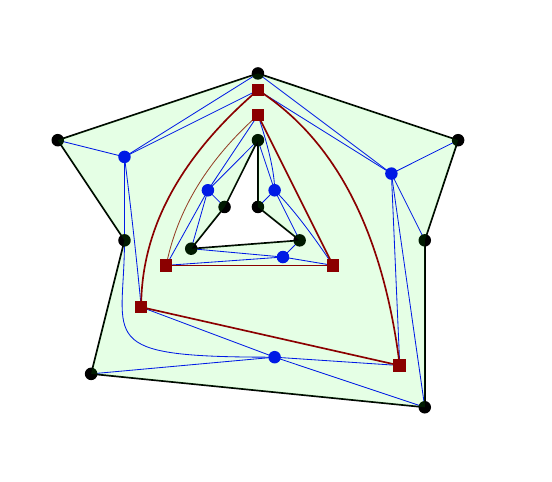}
    \includegraphics[width=0.4\textwidth, page=3]{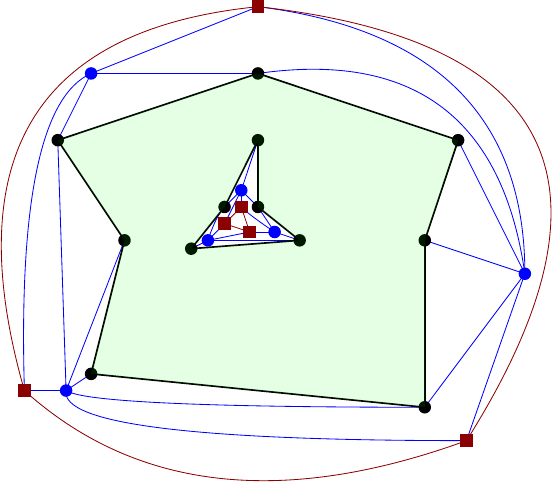}
    \caption{
    %\textbf{(Left)} old polygonal region $P$.
    \textbf{(Left)} Inside scaffolding $S^{in}(P)$.
    \textbf{(Right)} Outside scaffolding $S^{out}(P)$.
    }
    \label{fig:scaffold_example}
\end{figure}
Let $P$ be a $h$-holed polygonal region. We will want to map the boundary of $P$ to an $h$-holed polygonal region $Q$ which has a much simpler boundary. In particular, the boundary of each hole of $Q$ will be a triangle.
To do so, we will define a graph $G^{in}_S(P)$ (or $G^{out}_S(P)$) as the \emph{inside (outside) scaffold graph of $P$}, a graph where we place triangles inside (outside) of $P$ for each boundary hole of $P$.
In addition, we will define a mapping between the boundaries so we can apply \Cref{lem:boundary_routing} to route points between the boundaries. We will see uses for outside scaffold graphs in redrawing a graph in \Cref{sec:graph_redrawing} and inside scaffold graphs for constructing scaffold graphs of cutting-divisions in \Cref{sec:gluing}.

\begin{figure}
    \centering
    \includegraphics[width=0.4\textwidth, page=4]{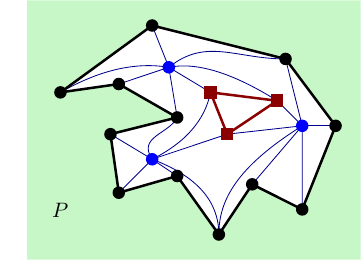}
    \hspace{1cm}
    \includegraphics[width=0.4\textwidth, page=5]{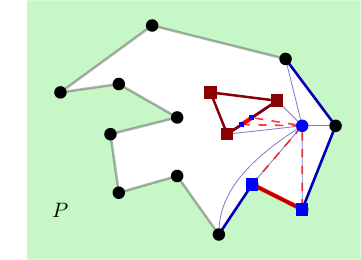}
    \caption{\textbf{(Left)} Outside scaffold graph for one hole. 
    \textbf{(Right)} The darker blue boundary of the outer polygon maps to the edge of the inner triangle. The red edge of the outer polygon maps to one part of the boundary of the triangle.}
    \label{fig:scaffold_polygon}
\end{figure}

In the subsequent section, we will define the construction of $S^{out}(P)$. The construction of $S^{in}(P)$ is similar, except we put points and edges on the interior of the polygon instead of the exterior. See \Cref{fig:scaffold_example} for an example of this.
For hole $i=1,2, \ldots, h$ of $P$ we will define edges and vertices of a graph:
\begin{itemize}
    \item Let $V_P^{(i)}$ be a collection of vertices defined by $i$th hole, and $E_p^{(i)}$ the edges connecting boundary of $V_P^{(i)}$.
    \item place three vertices $V_H^{(i)}$ on the exterior of the $i$th hole (the exterior of the outer hole is the outside), and join in a triangle by three edges $E_H^{i}$. 
    This 3-cycle of vertices $(V_H^{(i)}, E_H^{(i)})$ we will call a \emph{hole triangle}. 
    \item between the boundary of $P$ and the triangle formed by $(V_H^{(i)}, E_H^{(i)})$ we will place another three vertices in a triangle 
    $(V_O^{(i)}, E_O^{(i)})$
    inside the hole (or on the outside of $P$ for the outer hole).
    $V_O^{(i)}$ will be \emph{portal vertices}.
    \item
    Partition the boundary edges $E_P^{(i)}$ into three non-empty contiguous parts (arbitrarily), and map each contiguous part to one side of the $i$th hole triangle. 
    We add a set of scaffolding edges $E_S^{(i)}$ between all boundary vertices of one part of the partition, one of the portal vertices of hole $i$, and the two endpoints of the hole triangle that we mapped to such that everything remains planar. See \Cref{fig:scaffold_polygon} for an illustration of this for one hole.
\end{itemize}
Letting $V(S^{out}(P)) = \bigcup_{i=1}^h (V_P^{(i)} \cup V_H^{(i)}\cup V_O^{(i)})$ and 
$E(S^{out}(P)) = \bigcup_{i=1}^h (E_P^{(i)} \cup E_H^{(i)}\cup E_O^{(i)} \cup E_S^{(i)})$, we see that $S^{out}(P)$ is planar, and there is a mapping from every old boundary edge to some hole triangle edge of $Q = \bigcup_{i=1}^h E_H^{(i)}$. If there are $k$ boundary edges mapping to the same hole triangle edge, as the boundary edges must be contiguous, we can subdivide the hole triangle edge into $k$ equal parts and map each boundary edge into one of the subdivisions. See \Cref{fig:scaffold_polygon} for an illustration of this. 

\subsection{Boundary routing of nested polygons}
\label{sec:boundary_routing}
Next, we will prove a lemma that provides a method for routing polygonal paths between two given polygons, where the path is guaranteed to avoid intersections with other paths and consists of straight-line segments with few bends.

Consider a polygon region with one hole, where the outer boundary is a polygon $P$ and the inside boundary is a polygon $Q$.
Furthermore, consider if we identified every point on the boundary of $P$ with a different point on the boundary of $Q$ in a circular fashion. 
Formally that means we have a continuous homeomorphism $f: \partial(P) \to \partial(Q)$.
Suppose for a partition of $\partial(P)$ into $s$ intervals at the points $p_1, ..., p_s$, and we had a set of $s$ vertices $L$ of $s$ points $x_1, ..., x_s$ lying strictly in the interior of $P\setminus Q$ such that the $x_i$ can see all points of $\partial(P)$ between $p_i$ and $p_{i+1}$ as well as all points of $\partial(Q)$ between $f(p_i)$ and $f(p_{i+1})$ (where we take $p_{s+1} = p_1$). 
We call $L$ a set of \emph{partition respecting portal vertices}. 
Note that when we say that $x_i$ sees $p_i$, we mean that the straight line between $x_i$ and $p_i$ does not cross the boundary of $P$ or $Q$.
We can show the following lemma:

 \begin{lemma}[Boundary routing lemma]\label{lem:boundary_routing}
 Given two nested polygons $P$ and $Q$ and a homeomorphism $f:\partial(P) \to \partial(Q)$ identifying each point on $\partial(P)$ to $\partial(Q)$, and the boundary of $P$ partitioned at $p_1, ..., p_s$ and a set of partition respecting portal vertices $L$, 
 then for every point $p\in \partial(P)$, we can find a path from $p$ to $f(p)$ that can be drawn with straight lines with at most $3$ bends. 
 Furthermore, for any two paths corresponding to different points on $\partial(P)$, the paths do not cross.
 \end{lemma}
 \begin{proof}
 
\begin{figure}
    \centering
    \includegraphics[height=0.2\textheight]{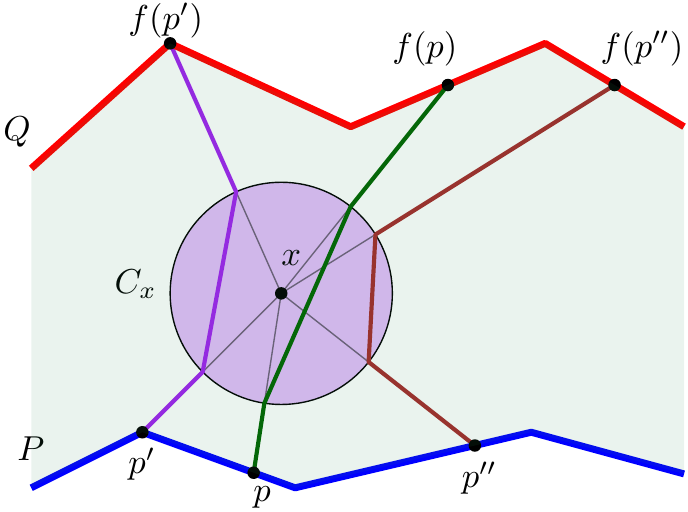}
    \hspace{4em}
    \includegraphics[height=0.2\textheight,page=3]{polygon_routing2.pdf}
    \caption{
    \textbf{(Left)} Routing a point $p$, $p'$, and $p''$ through the portal $x$ via circle $C_x$.
    \textbf{(Right)} Routing a point $p$, $p'$, and $p''$ through the portal $x$ via triangle $T_x$.}
    \label{fig:polygon_routing}
\end{figure}
 We will describe how we draw paths from points on the boundary of $P$ between $p_1$ and $p_2$. This section of the boundary has a corresponding portal $x=x_1\in L$.
 Let $p$ and $p'$ and $p''$ be any three different points on the boundary of $P$ between $p_1$ and $p_2$ such that $p$ lies between $p'$ and $p''$.
 Note that we would like to choose the path that is the straight line from $p$ to $x$ followed by the straight line from $x$ to $f(p)$ which would have at most one bend, and similarly for $p'$.
However, if we did that, the paths for $p$, $p'$, and $p''$ would intersect at $x$.

Instead, let us consider a circle $C_x$ of radius $\eps$ around $x$ for some sufficiently small radius $\eps>0$ that does not contain any point of the boundary of $P$ or $Q$. Note that we can always find an $\eps$ as $x_1$ lies in the interior of $P\setminus Q$.
Let $C_x(y)$ denote the point of intersection between $C_x$ and the line from $x$ to $y$.
We will instead let the path connecting $p$ to $f(p)$ to be the straight line from $p$ to $C_x(p)$ to $C_x(f(p))$ to $f(p)$ as in \Cref{fig:polygon_routing}.
Note that by the homeomorphism $f$, the paths defined by $p'$ and $p''$ will not intersect the one for $p$.
 \end{proof}

\paragraph{Remark.} Note that computing the intersection of a line with a circle involves computing a square root. 
It is possible to avoid this by replacing the circle $C_x$ with a carefully chosen triangle $T_x$ so that no point $p$ and $f(p)$ lie on the same side of the triangle.
Consider if we are trying to map the section along the boundary $p_1$ to $p_2$ to the section $f(p_1)$ to $f(p_2)$.
Note that the lines formed by straight lines from the boundary sections of at least one of them must define an angle of at most $180^\circ$,
without loss of generality let that be the boundary section from $f(p_1)$ to $f(p_2)$. 
Let $T_x$ denote the triangle formed by a line parallel to the line between $f(p_1)$ and $f(p_2)$ of distance $\eps \approx 1/N^{O(1)}$ away from $x$ (where $N$ is the total number of points of $P$ and $Q$) and a third point at a distance of $\eps$ in the other direction as depicted in \Cref{fig:polygon_routing}. If all input points were rational coordinates representable with $O(\log N)$ bit numerators and denominators, it is clear that this polygon has rational coordinates that are representable in $O(\log N)$ bits, and all intersections with the boundary of this triangle also have rational coordinates representable in $O(\log N)$ bits.

\subsection{Redrawing a polygonal embedding as a triangular polygonal embedding}
\label{sec:graph_redrawing}
As an application of our scaffold graph, we will show in the following theorem that we can redraw any $(G, \nabla(G), P)$ to a polygonal domain $Q$ where the boundary of each hole of $Q$ is a triangle. 
\begin{restatable}[Redrawing lemma]{lemma}{redrawing}
\label{lem:redrawing}
Given an $h$-holed embedding $(G, \nabla(G), P)$
it is possible to find  a %n aligned 
polygonal embedding $(H, \nabla(G), Q)$ where $G$ is a minor of $H$,
$Q$ is a triangular polygon,
all vertices of $\nabla(G)$ are on the boundaries of $Q$,
and all edges are drawn in an $O(n)\times O(n)$ grid as a straight line lying in $Q$ with $O(h^2)$. 

We can do this even if we are only given a combinatorial description of $(G, \nabla(G), P)$ instead of an explicit embedding.
\end{restatable}
\begin{proof}
We may assume $P$ is a polygon with vertices at every vertex of $\nabla(G)$ (if not, we may subdivide edges on the boundary of $P$).
We begin by constructing the outside scaffold graph $S^{out}(P)$. Let $G' = S^{out}(P) \cup G$ be the union of the scaffolding graph of the polygon and our planar graph. Note that this graph is planar.
Use a known algorithm to draw $G'$ with the special $3$ outer vertices of $S^{out}(P)$
in an $O(n)\times O(n)$ grid.
This can be done using \Cref{thm:graph_drawing}.

Now, this drawing induces a polygonal region $Q$ with holes defined by the triangles for each hole. We can view $G'$ as a graph drawn on $Q$.
Let's consider the $i$th hole of $P$. There is a hole triangle defined by $(V_H^{(i)}, E_H^{(i)})$ 
Consider a vertex $x\in V_P^{(i)}$ corresponding to a vertex of $\nabla(G)$, that should be at a location $x'$ on some edge of $E_H^{(i)}$. 
By the Boundary Routing Lemma of \Cref{lem:boundary_routing}, for each such $x$, we can draw a polyline $e_x$ between $x$ and $x'$ with at most $O(1)$ bends such that none of the polylines for any vertex cross by routing through the corresponding portal vertex.
Let this resulting graph be $H$. By contracting these bends and other auxiliary edges added in the scaffold graph, it is clear that $H$ contains $G$ as a minor.

Note that we only used local features of $(G, \nabla(G), P)$ to construct the scaffolding graph. It suffices if we are given a combinatorial description of a planar embedding $(G, \nabla(G), P)$.
\end{proof}

\subsection{Gluing planar graphs together} \label{sec:gluing}

Suppose we are given a polygonal embedding $(G, \nabla(G), P)$ where $G$ has $n$ edges and $P$ has $O(1)$ holes and an $(s, 1/r)$-cutting-division $\Gamma$ of $(G, \nabla(G), P)$. 
Let $(G_i, \nabla(G_i), P_i)$ be the induced graphs in each region $P_i\in \Gamma$.

Suppose we want to change each $G_i$ to some graph $H_i$ and reduce the complexity of each polygonal region from $P_i$ to $Q_i$, a polygonal region with the same number of holes and each hole with $3$ sides. 
Afterward, we would want to join these graphs $(H_i, \nabla(G), Q_i)$ into a single graph embedded graph $(H, \nabla(G), Q)$ where $Q$ is a polygonal region that has the same number of holes as $P$ with each side a triangle and for every vertex of $\bigcup_{i=1}^k \nabla(G_i)$ that correspond to the same point in multiple boundaries are joined by an edge with $O(1)$ bends. We will call $H$ the \emph{glued graph}, and $(H, \nabla(G), Q)$ the $\emph{glued embedding}$ of $(H_i, \nabla(G_i), Q_i)$.
Formally we state the lemma we aim to prove.

\begin{restatable}[Gluing Lemma]{lemma}{gluinglemma}
\label{lem:gluing_lemma}
Suppose we have an $O(1)$-holed embedded planar graph $(G, \nabla(G), P)$ and an $(s,1/r)$-cutting-division $\Gamma$ that induces polygonal embeddings $(G_i, \nabla(G_i), P_i)$. Suppose for every $i$ we are given another polygonal embedded graph with the same boundary vertices $(H_i, \nabla(G_i), Q_i)$ where $Q_i$ is a triangular polygon with the same number of holes as $P_i$.
Then we can \emph{glue} together the $(H_i, \nabla(G_i), Q_i)$ to form a polygonal embedding with 
$(H, \nabla(G), Q)$ where $|E(H)| = O(|\nabla(G)| + \sum_{i=1}^k |H_i|)$ and $Q$ is a triangular polygon with the same number of holes as $P$.

Furthermore, we can do this in the MPC model of computation in $O(1)$ rounds with $O(|E(G)|/\cS)$ machines where each machine uses $O(\cS)$ storage as long as $|\Gamma| \le \cS^{1-\Omega(1)}$.
\end{restatable}

We briefly describe why this is not an easy task to do in the MPC model. 
Our cutting-division $\Gamma$ has size roughly $n^{\delta}$ for some constant $\delta>0$, and can have arbitrarily complicated topology, but we need to do the gluing in $O(1)$ rounds.
For example, we may have polygonal regions nested (i.e., completely contained) in another polygonal region, but also adjacent to another nested polygonal region.
The polygonal embeddings we need to glue together have triangular boundaries, it is not obvious how to connect the boundaries up together.
For nested polygonal embeddings, they may be arbitrarily deeply nested, so we cannot use recursion to draw each nested region, since that would take $O(n^{\delta})$ rounds, instead, we need a method of drawing the entire graph in one go. 
Another issue is that we need to compute precise coordinates for every point we are gluing together, so we have to be careful to not increase the bit complexity of our drawings by too much.
One last issue that we will need to be concerned about is that we can also have many polygonal regions that all share one single vertex.
We do not want to add too many extra edges or bends, and we can only add a number proportional to the number of points on the boundary.

Thus, it is necessary to do the entire gluing in one single step so that we only take $O(1)$ rounds of communication. 
To do so we will use a \emph{scaffold graph for cutting-divisions} that we will define below.
At a high level, the scaffold graph provides us a place for us to put each of the triangular embeddings $Q_i$ (hence it serves as a scaffold) and gives a way for us to connect the boundaries of the $Q_i$ to the boundary of $\Gamma$ in the scaffold graph. This resolves the issue of a boundary vertex that is on many polygonal regions, as there will be one copy of the vertex on the scaffold graph that each polygonal region will route to independently.

\paragraph{Scaffold graph for cutting-divisions.}
%\label{sec:cutting-division_scaffold}
Let $\Gamma$ be an $(s, 1/r)$-cutting-division of an $O(1)$-hole embedded planar graph $(G, \nabla(G), P)$.
Recall that $\Gamma$ partitions $P$ into $k = O(r/s)$ disjoint regions defined by polygonal regions $P_1,..., P_k$. We define the \emph{scaffold graph of $\Gamma$} to be a graph $S(\Gamma)$ with inside scaffolding for each polygonal region $P_i$ and outside scaffolding for each hole of $P$. 
Formally, we let $S(\Gamma)$ be the graph with vertices at $P_i$ and edges defined as the edges of each $P_i$, together with the inside scaffold $S^{in}(P_i)$ for each $i=1,...,k$ as well as $S^{out}(P)$.

\begin{proof}[Proof of \Cref{lem:gluing_lemma}]
    Construct the scaffold graph for a cutting-division $S(\Gamma)$.
    We can compute a straight-line drawing of the scaffold graph $S(\Gamma)$ by \Cref{thm:graph_drawing} in an $N\times N$ grid where $N = O(|\Gamma| + |\nabla(P)|)$.
    For each $(H_i, \nabla(G_i), Q_i)$, we can embed in the corresponding polygonal region $Q_i'\in S(\Gamma)$ by the Triangular Morphing Lemma of \Cref{lem:morph}.
    Then we can apply the Boundary Routing Lemma (\Cref{lem:boundary_routing}) to connect vertices to the old boundary edges of $P_1, ..., P_k$. 
    Furthermore, observe that each application of the Boundary Routing Lemma only increases the number of bends per edge by $O(1)$. The Triangular Morphing Lemma increases the number of bends by $O(1)$ as each region has $O(1)$ holes. 

    To show that we can do this in the MPC model of computation, it suffices to show that we can apply the Boundary Routing Lemma and the Triangular Morphing Lemma in the MPC model. First, observe that as $|\Gamma| \le \cS^{1-\Omega(1)}$, the cutting-division can be distributed to all machines.
    The Triangular Morphing Lemma is simple to implement in the MPC model since $Q_i$ and $Q_i'$ are triangular polygonal regions with $O(1)$ holes and thus the linear homeomorphism between the two have $O(1)$ complexity and thus can be computed locally whenever necessary and applied to each edge.
    On the other hand, the Boundary Routing Lemma may need to be applied to many edges. However, the routing is deterministic and can be computed for each machine independently. 
    In particular, we can deterministically and in parallel compute the locations of vertices on the boundary of $\Gamma$.
    Thus each machine can compute how to route the edges that are stored on it, and the Boundary Routing Lemma will guarantee that no two edges cross.
\end{proof}

\subsection{Handling degenerate polygonal regions} \label{sec:degenerate}
\begin{figure}
    \centering
    \includegraphics[width=0.4\textwidth, page=5]{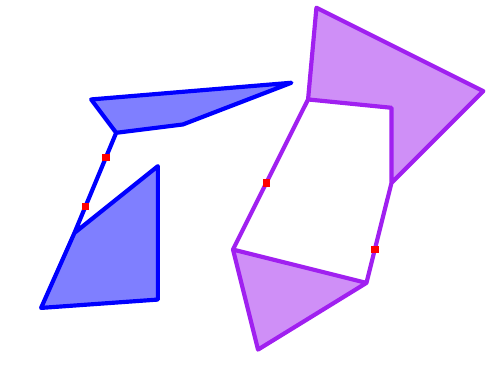}
    \includegraphics[width=0.4\textwidth, page=7]{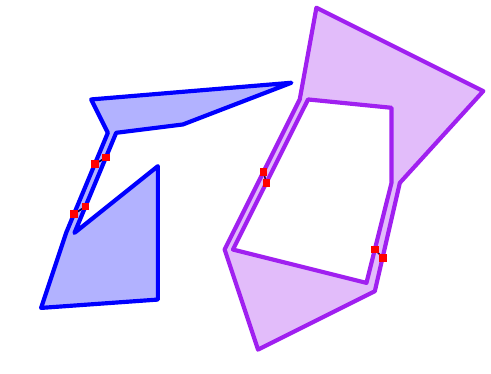}
    \caption{For degenerate polygons with boundary vertices we can imagine the degenerate regions being slightly separated and duplicate boundary vertices on each side being connected by an edge.}
    \label{fig:degenerate_polygons}
\end{figure}

To handle degenerate input polygons, we will describe a way to remove the degeneracy at a small cost of adding a few more edges into our graph (proportional to the number of boundary vertices). 
This has the added benefit of guaranteeing that the output polygonal region of an embedding has no degeneracies.

For every degenerate section of the polygonal region, we can treat them as very slightly to treat it as a non-degenerate polygon.
There may be boundary vertices of $\nabla(G)$ on these degenerate regions.
For these, we will duplicate the boundary vertices so they are on both boundaries and add an edge connecting them, and treat this as a vertex split. 
Note that in reality, these duplicate boundary vertices will be at the same location, and this edge joining the two boundaries has zero length.
This is illustrated in \Cref{fig:degenerate_polygons} where we have slightly expanded the degenerate regions for illustrative purposes.  
We can do this as soon as we detect that we have a degenerate section of the boundary as we construct degenerate polygons in the cutting-division. 
These extra degenerate edges can be constructed in parallel, as all machines are aware of the cutting-division.

\section{MPC algorithms for embedded planar graphs}\label{sec:algorithms}

We show in this section how to construct $O(1)$ round algorithms for fundamental problems on embedded planar graphs using $(s, 1/r)$-cutting-divisions. 
The high-level plan is to use cutting-divisions to break our problem into smaller parts so we can recurse on each part. 
We recurse until the problem is small enough to be solved on one machine.
There is often part of the problem that remains unsolved involving boundary vertices because the problems involve the rest of the graph.
Thus we will ``compress'' the graph to have size roughly on the order of the number of boundary vertices.
We will end we can glue together the compressed graphs to get a slightly smaller graph, so we can use recursion.
The exact details of how we do this vary depending on the specific problem at hand.

To begin, we will illustrate how we can redraw a polygonal embedded graph so that the boundary is a triangular polygonal region. Recall that triangular polygonal regions have all holes being triangles so such a triangular polygonal region has $O(1)$ boundary complexity.
This is a useful subroutine as when we glue graphs together in $O(1)$ rounds, we need to have embeddings in triangular polygonal regions.
Afterward, we will describe algorithms for connected component and minimum spanning forest.
Next, we describe algorithms for $(1+\eps)$-approximate path type problems that require computing $\eps$-emulators.
After this, we describe algorithms for finding $st$-shortest paths, shortest cycle, single source shortest path (SSSP), and all-pairs shortest path (APSP).
Finally, we will describe how we can draw a primal-dual overlay graph, a graph containing both the old primal graph and the dual graph.
Using this, we will show how we can compute $st$-max-flows and min-cuts.

We note that we will cover the most fundamental problems that we hope is enough to illustrate the power of our recursive framework.
There are many more problems and variations of problems we will refrain from discussing, like how to label vertices of each connected component with the same label or computing a bipartition, which are straightforward applications of our framework.
For the $(1+\eps)$-approximate path type problems like shortest paths, cycles, and flows, we can also recover the answer by running our algorithm backward from the final compressed graph.

Throughout this section, we will use $n =|V(G)|$ to denote the number of vertices of the graph, and $m=|E(G)|$ to denote the number of edges of the graph. Note that for planar graphs without isolated vertices $m = \Theta(n)$, so we sometimes use $m$ and $n$ interchangeably. Isolated vertices can easily be detected and handled depending on the problem we are aiming to solve.

\subsection{Redrawing a graph}
As a warmup, we will illustrate one of the more simple consequences of the gluing lemma to redraw an $O(1)$-holed polygonal embedded graph $(G, \nabla(G), P)$ into a triangular polygonal region $Q$ with the same number of holes as $P$.
In this proof, we will carefully analyze the bit complexity of the algorithm, and discuss the cyclic ordering of vertices along the boundaries of each hole. In later proofs, these details were omitted for the sake of brevity. Bit complexity is important when working in the MPC model because each machine is bounded to $O(\cS)$ words of space, if the bit complexity blows up, we may not be able to do recursion without using more total space. 

\begin{lemma}[MPC Redrawing Lemma] \label{lem:mpc_redrawing}
Let $r = \cS^\alpha$ for a sufficiently small constant $\alpha > 0$.
There exists an algorithm that 
takes as input a $O(1)$-holed polygonal embedded graph $(G, \nabla(G), P)$ with $n$ vertices and $m$ edges with $|\partial(P)| = O(r^{1/3})$
and returns a polygonal embedded graph $(H, \nabla (G), Q)$ 
satisfying these properties:
\begin{itemize}
    \item $Q$ is a triangular polygonal region with the same number of holes as $P$
    \item $V(H) \supseteq V(G)$
    \item $H$ contains $G$ as a minor
    \item Each edge $e \in E(H)$ is drawn with a straight line
    \item $|E(H)| = O(m)$ 
    \item The coordinates of $V(H)$ and $Q$ are stored as rationals with $O(\log n)$ bit complexity.
    \item The cyclic ordering of the vertices of $\nabla(G)$ around each hole in $(H, \nabla(G), Q)$ is the same as in $(G, \nabla(G), P)$.
\end{itemize}
The algorithm runs in $O(1)$ rounds in expectation and with high probability 
using $\Theta(\cS)$ space per machine and $O(n/\cS)$ machines where $\cS = n^{\delta}$ for any constant $\delta > 0$.
\end{lemma}

\begin{proof}
We give a recursive algorithm.
\paragraph{Base case.} If $m \le  \cS$, we can solve the problem directly on one machine using \Cref{lem:redrawing}. 
The drawing output by the lemma has vertices stored as rationals with $O(\log m)$ bit complexity.

\paragraph{Recursive case.} Otherwise, if $|G| > \cS$, we apply \Cref{lem:mpc_cutting-division} to construct a $(r^{2/3}, 1/r)$-cutting-division.
This decomposes the problem into $k=O(r^{1/3})$ polygonal embeddings $(G_1, \nabla(G_1), P_1), ..., (G_k, \nabla(G_k), P_k)$ where for each $1\le i \le k$, $|G_i| \le O(m/r^{1/3})$, $|\partial(P_i)| = O(|\partial(P)| + \sqrt{r^{2/3}}) = O(r^{1/3})$.

Note that as we have $O(r^{1/3})$ problems each with $O(n/r^{1/3})$ edges and total boundary size $O(\cS^{1/3})$.
We can solve all the recursive subproblems in $O(1)$ rounds in parallel by recursion after partitioning edges and subdividing edges to get graphs $(H_i, \nabla(G_i), Q_i)$
where each graph has coordinates as rationals with $O(\log m)$ bit complexity and the cyclic ordering around each hole matches that in $(G_i, \nabla(G_i), P_i)$.
Thus we can apply the \Cref{lem:gluing_lemma} to glue together $(H_i, \nabla(G_i), Q_i)$ from the recursive calls to get a single graph $(H, \nabla(G), Q)$.
Note that the gluing lemma will apply piecewise linear transformations to each $(H_i, \nabla(G_i), Q_i)$, but these transformations can be expressed as linear functions on the coordinates with coefficients as rationals with $O(\log m)$ bit complexity, so the total bit complexity of any coordinate does not increase by more than a constant.

It remains to analyze the number of rounds this algorithm takes.
Let $T_{draw}(m)$ denote the number of rounds for an $m$ edge polygonal embedded graph.
Then it is clear that this algorithm follows the recurrence:
\[ T_{draw}(m) = T_{draw}(m/r^{1/3}) + O(1) \]
Since $r = \cS^{\alpha} = n^{\alpha \delta} = \Theta(m^{\alpha \delta})$ the recurrence solves to $T_{draw}(m) = O(1/(\alpha \delta)) = O(1)$.
\end{proof}

\subsection{Connected components}
Let $\cc(G)$ denote the number of components in the graph.

\thmCC*

To prove this theorem, we prove the following claim.
\begin{claim} \label{clm:cc}
Let $r = \cS^\alpha$ for a sufficiently small constant $\alpha > 0$.
There exists an algorithm that 
given a $O(1)$-holed polygonal embedded graph $(G, \nabla(G), P)$ with $m=|G|$ edges 
where $|\partial(P)|= O(r^{1/3})$,
returns an embedded graph $(H, \nabla(G), Q)$ and an integer $\ell$ satisfying these properties:
\begin{itemize}
    \item $Q$ is a triangular polygonal region
    \item $V(H) \supseteq \nabla(G)$
    \item $|H| < c_0 |\nabla(G)|$ for some constant $c_0$ 
    \item $\cc(G) = \cc(H) + \ell$ 
    \item two vertices in $\nabla(G)$ are connected in $H$ iff they are connected in $G$. 
\end{itemize}
The algorithm runs in $O(1)$ rounds 
%in expectation and with high probability 
using $\Theta(\cS)$ space per machine and $O(n/\cS)$ machines where $\cS = m^{\delta}$ for any constant $\delta > 0$.
\end{claim}

\begin{proof}
We give a recursive algorithm. 

\paragraph{Base case.} 

If $m \le \cS$, we can solve this problem directly on one machine.
To do so, we can compute $\ell$ the number of connected components not connected to any boundary vertices, and find a collection of disjoint paths on $\nabla(G)$ that has the same connectivity as $\nabla(G)$. Concretely, this can be done by computing a minimum spanning forest on $\nabla(G)$, then letting the path be the sequence of vertices of $\nabla(G)$ visited on an Euler tour of the MST for each component.
This collection of disjoint paths is clearly planar and can be drawn in the polygon $Q$ with $O(1)$ bends per edge. 

\paragraph{Recursive case.} 
If $m < c_0 |\nabla(G)|$ then we can simply redraw $G$ with \Cref{lem:mpc_redrawing}, and return $H$ as the redrawn $G$, and $\ell = 0$. 

Otherwise $m \ge c_0 |\nabla(G)|$. 
As in \Cref{lem:mpc_redrawing} we construct a $(r^{2/3}, 1/r)$-cutting-division.
This splits the problem into subgraphs $G_1,..., G_k$ for $k=O(r^{1/3})$.
We can apply recursion to each subproblem to get $H_i$ and $\ell_i$ for each subproblem.
Let $\ell = \sum_{i=1}^k \ell_i$. 
We can use the Gluing Lemma, \Cref{lem:gluing_lemma}, to draw each $H_i$ in a triangular polygonal region $Q_i$ and glue the polygonal embeddings together to get $(H, \nabla(G), Q)$. 

We will recurse on $H$. To show that our algorithm terminates, we need to ensure that $|E(H)| < m$.
Suppose that the Gluing Lemma adds at most $c_1$ bends to each edge. 
Then:
\[|E(H)| \le c_1\left(|\nabla(G)| + O\left(\sum_{i=1}^k |\nabla(G_i)|\right)\right) = c_1|\nabla(G)| + O(m/r^{1/3}) \]
We can assume $O(m/r^{1/3}) \le m/3$ since $r = \Theta(m^{\alpha \delta})$.
Thus if $c_0= c_1/3$ it follows that $|E(H)| \le 2m/3 < m$.
This means we can recurse on the polygonal embedding $(H, \nabla(G), Q)$ to get a polygonal embedding $(H', \nabla(G), Q')$ with $|H'| = O(|\nabla(G)|)$ and some $\ell'$. Now we can simply return $\ell + \ell'$ along with this embedding.

It remains to argue that this algorithm terminates within $O(1)$ rounds. Let $T_{cc}(n)$ denote the total number of rounds it takes to solve the problem in the claim with $n$ edges. The number of rounds that this algorithm takes is given by this recursion.
\begin{align*} T_{cc}(m)  &= \max\left(T_{draw}(m), T_{cc}(|H|) +\max_{i=1,...,k} T_{cc}(|G_i|) + O(1) \right)  
\\&= \max\left(T_{draw}(m), T_{cc}(m/r^{1/3}) + O(1)\right) \end{align*}
 Since $T_{draw}(m) = O(1)$ and $r = \Theta(m^{\alpha \delta})$, this recursion solves to $T_{cc}(m) = O(1)$.
\end{proof}

\paragraph{Remark.} Instead of counting, we can put a label on each edge and vertex such that every connected component has a different label, and two edges/vertices have the same label iff they are in the same connected component. This is easy to do in the base case on one machine. To do this for the recursive step, after we get connected components of the compressed graph, we need to modify the labels of the uncompressed components. This can be done in parallel in $O(1)$ rounds by ``uncompressing'' the graph. Formally proving this is somewhat cumbersome, so we will omit the proof for brevity.

\subsection{Minimum spanning forest}

Without loss of generality, we may assume that the edge weights of the tree are unique by breaking ties by the unique identifiers of edges.
This means the minimum spanning forest is unique.
Furthermore, we can assume all edge weights are positive (if not, we can add a large positive number to each edge).

\thmMST*

%We can also compute the weight of the minimum spanning forest by summing the weights over all marked edges in an additional $O(1)$ rounds.

To prove the theorem we use the following lemma about minimum spanning trees. 
The following lemma is folklore.
\begin{lemma} \label{lem:mst_contraction}
Let $G$ be a connected graph and let $T_G$ be the minimum spanning tree for $G$. Let $e$ be an edge in $T_G$. Let $G'$ be $G$ after contracting $e$ and $T_{G'}$ be the minimum spanning tree for $G'$.  
Then $T_G = T_G' \cup \{e\}$.
\end{lemma}

We now outline the algorithm. We will apply our cutting-division and recurse on each component until the problem size is small enough, as we did for counting connected components.
To solve small-sized problems we will compute a minimum spanning forest for the small instance and use the above lemma to contract all vertices that do not lie on the boundary of the problem. As planar graphs remain planar upon contraction, we will have a much smaller planar graph with number of edges proportional to the number of boundary vertices.
This reduces the size of the graph so we can apply recursion.

To prove \Cref{thm:mst}, we prove the following claim. \Cref{thm:mst} follows, as if we let $B$ be a large enough square, then we can apply the claim to $(G, \emptyset, B)$ to get a minimum spanning forest.
\begin{claim}
Let $r= \cS^{\alpha}$ for a sufficiently small constant $\alpha$.
There exists an algorithm that 
given an $O(1)$-holed polygonal embedded graph $(G, \nabla(G), P)$ with $m$ edges and $n$ vertices 
where $|\partial(P)|=O(r^{1/3})$,
returns an embedded graph $(H, \nabla(G), Q)$ 
and a set $E_T\subseteq E(G)$ of edges satisfying:
\begin{itemize}
    \item $Q$ is a triangular polygonal region 
    \item $V(H) \supseteq \nabla(G)$
    \item $|H| < c_0 |\nabla(G)|$ for some constant $c_0$ 
    \item Let $E_G$ denote the edges of the minimum spanning forest of $G$. Let $E_H$ denote the canonical edges of $G$ in the minimum spanning forest of $H$.
    Then $E_G = E_T \cup E_H$.
\end{itemize}
The algorithm runs in $O(1)$ rounds %in expectation and with high probability 
using $\Theta(\cS)$ space per machine and $O(n/\cS)$ machines where $\cS = n^{\delta}$ for any constant $\delta > 0$.
\end{claim}
\begin{proof}
Note that throughout our algorithm our graph will subdivide edges and split vertices apart and connect them by an edge. 
We will view these added edges (and vertices) as \emph{virtual edges} (and \emph{virtual vertices}). 
They will have weight $0$ (so will be added into a minimum spanning forest at $0$ cost.  On vertex splits, all added edges will be virtual.
On subdivision of a non-virtual edge $e$ into two edges $e_1$ and $e_2$, we will arbitrarily let $e_2$ be a virtual edge, and let $e_1$ be the canonical edge storing information about the old edge in $G$.
Observe that any minimum spanning forest will use the old subdivided edge $e$, if and only if it uses both $e_1$ and $e_2$. Since $e_2$ is a virtual edge with weight $0$, we will always include it in our minimum spanning forest. Thus this does not impact the correctness of our solution, we will take the canonical edge $e_1$ for the minimum spanning forest of the graph after subdivision if and only if we took $e$ for the minimum spanning forest in the old graph.

\paragraph{Base case.} If $m\le \cS$, we can solve this problem directly on one machine. For any component that is not connected to $\nabla(G)$ we may use any minimum spanning tree algorithm on that component.
Compute a minimum spanning tree $T$ on every other component. Repeatedly contract edges that do not join two vertices of $\nabla(G)$ which we can do by \Cref{lem:mst_contraction} and add the contracted edge to $E_T$ if it were a canonical edge.  
Note that planar graphs remain planar on edge contraction (though we may no longer have an embedding). 
Computing the minimum spanning tree can be done in $O(m\log m)$ time.
Afterward, we are left with a planar graph $H$ on the vertices of $\nabla(G)$ that we can embed in a triangular polygonal region $Q$ by the \Cref{lem:redrawing}.

\paragraph{Recursive case.} The recursive case is exactly the same as for our algorithm for connected components of \Cref{clm:cc} except instead of returning $\ell$, we mark edges of $G$ as in $E_T$.
\end{proof}

\subsection{Computing emulators}

To solve $(1+\eps)$-approximate problems involving shortest paths we will construct $\eps$-emulators in the MPC model.
The algorithm is very similar to our algorithm for redrawing the graph.
The main difference is in the base case, we will instead construct the $\eps$-emulator using the algorithm of Chang, Krauthgamer, and Tan \cite{ChangKT22}.
However, if the graph we have already has size roughly equal to the size of the boundary (up to polylogarithmic factors), we do not need to recurse, and instead it suffices to redraw the graph, which we know we can do in $O(1)$ rounds.
One might wonder why we need to redraw the graph at all. The reason is that we require the boundary polygon for each region to have $O(1)$ boundary for the Gluing Lemma.

\begin{lemma}[MPC $(1+\eps)$-emulator] \label{lem:mpc_emulator}
Let $r= \cS^{\alpha}$ for a sufficiently small constant $\alpha$.
There exists an algorithm that given 
as input 
an $O(1)$-holed polygonal embedded graph $(G, \nabla(G), P)$ with $m$ edges and $n$ vertices 
where $|\partial(P)|=O(r^{1/3})$,
returns an embedded graph $(H, \nabla(G), Q)$ satisfying:
\begin{itemize}
    \item $Q$ is a triangular polygonal region 
    \item $V(H) \supseteq \nabla(G)$
    \item $(H, \nabla(G))$ is an $\eps$-emulator of $(G, \nabla(G))$.
    \item $|H| < c_0 |\nabla(G)|\log^D(|\nabla(G)|)$ for some constant $c_0$ and integer $D$
\end{itemize}
The algorithm runs for $O(1)$ rounds in expectation and with high probability 
using $\Theta(\cS)$ space per machine and $O(n/\cS)$ machines where $\cS = n^{\delta}$ for any constant $\delta > 0$.
\end{lemma}

\begin{proof}
We present a recursive algorithm. The base case is a sequential algorithm. Throughout the algorithm, we subdivide edges and split vertices. 
Recall that when subdividing an edge $e$ into $e_1$ and $e_2$, we choose $e_1$ to be the canonical edge and let $w(e_1) = w(e)$ and $w(e_2) = 0$.
When splitting a vertex $v$ into $u_1$ and $u_2$ by an edge $e=(u_1, u_2)$, we set $w(e) = 0$.
Note that this does not change the distances between vertices (or copies of vertices).
This means that if $(G, \nabla(G))$ is an $\eps$-emulator, then for any graph $H$ constructed by subdividing edges and splitting vertices of $G$ in this manner, we will have that $(H, \nabla(G))$ is also an $\eps$-emuulator.

\paragraph{Base case.} We run the base case if $m \le \cS$.
Observe that the $\nabla(G)$ lies on topological holes of $(G,\nabla(G))$.
We compute an $\eps$-emulator of $(G, \nabla(G))$ using the algorithm of Chang, Krauthgamer, and Tan \cite{ChangKT22} stated in \Cref{thm:eps_emulator} to get a $(H, \nabla(G))$ with $H$ having the same number of topological holes as $G$ where $|E(H)| = O(|\nabla(G)|\polylog(|\nabla(G)|))$.
Using \Cref{lem:redrawing} we can compute $(H, \nabla(G), Q)$ for a triangular region $Q$. 

\paragraph{Recursive case.}
The recursive case follows almost exactly as in our previous recursive algorithms, the extra polylogarithmic factors in the size of the $\eps$-emulators do not make a difference. 
For completeness, we present the full proof.

If $m \le c_0 |\nabla(G)|\log^D(|\nabla(G)|)$, then we can redraw the graph $G$ as $H$ using \Cref{lem:mpc_redrawing}.
Note that $(H, \nabla (G))$ is an exact distance emulator of $(G, \nabla(G))$.

Otherwise $m \ge c_0 |\nabla(G)|\log^{D}(|\nabla(G)|)$. 
As in \Cref{lem:mpc_redrawing} we construct a $(r^{2/3}, 1/r)$-cutting-division.
%for  $r= \cS^{\alpha}$ for a sufficiently small constant $\alpha$.
This splits the problem into subgraphs $G_1,..., G_k$ for $k=O(r^{1/3})$.
We can apply recursion to each subproblem to get $H_i$ for each subproblem.
We can use the Gluing Lemma, \Cref{lem:gluing_lemma}, to draw each $H_i$ in a triangular polygonal region $Q_i$ and glue the polygonal embeddings together to get $(H, \nabla(G), Q)$. 

We will recurse on $H$. To show that our algorithm terminates, we need to ensure that $|E(H)| < m$.
Suppose that the Gluing Lemma adds at most $c_1$ bends to each edge.
Thus observe that if we choose $c_0 = 3c_1$:
\[|E(H)| \le c_1\left(|\nabla(G)| + O\biggl(\sum_{i=1}^k |\nabla(G_i)|\biggr)\right) \le c_1|\nabla(G)| + \OO\left(\frac{m}{r^{1/3}}\right) \le \frac{c_0}{3}|\nabla(G)| + \OO\left(\frac{m}{r^{1/3}}\right). \]
We can assume $\OO(m/r^{1/3}) \le m/3$ since $r = \Omega(m^{\alpha \delta})$.
Thus it follows that $|E(H)| \le 2m/3 < m$.
This means we can recurse on the polygonal embedding $(H, \nabla(G), Q)$ to get a polygonal embedding $(H', \nabla(G), Q')$ with $|H'| = O(|\nabla(G)|)$. 

It remains to argue that this algorithm terminates within $O(1)$ rounds. Let $T_{emu}(m)$ denote the total number of rounds it takes to solve the problem in the claim with $m$ edges. Clearly, the number of rounds that this algorithm takes follows the recursion:
\begin{align*}
T_{emu}(m) 
&= \max\left(T_{draw}(m), \max_{i=1,...,k} T_{emu}(|E(G_i)|) + T_{emu}(|E(H)|) + O(1) \right)  \\
& = \max\left(T_{draw}(m), T_{emu}\left(\OO\left(\frac{m}{r^{1/3}}\right)\right) + O(1)\right)
 \end{align*}
Since $T_{draw}(m) = O(1/\delta)$, this recursion solves to $T_{emu} = O(1)$ as $r = \Theta(m^{\alpha \delta})$.
\end{proof}

\paragraph{$st$-shortest paths.} The following \Cref{cor:st_shortest_paths} about $st$-shortest paths easily follows from \Cref{lem:mpc_emulator}. For a graph $G$ and vertices $s,t, \in V(G)$, consider a large enough bounding box with two punctures (small holes) at $s$ and $t$. Call this punctured box $B$. Apply \Cref{lem:mpc_emulator} to $(G, \{s, t\}, B)$ to get a graph $(H, \{s, t\}, Q')$ of size $\OO(1)$ such that $(H, \{s, t\})$ is a $(1+\eps)$-emulator. Here we can easily compute a shortest path between $s$ and $t$ in $H$ on a single machine however we'd like, and get a $(1+\eps)$-approximate shortest path between $s$ and $t$. 
\begin{corollary}[$(1+\eps)$-approximate $st$-shortest paths]
\label{cor:st_shortest_paths}
    Given an embedded planar graph $G$ with $m$ edges and $n$ vertices and two vertices $s, t\in V(G)$. There is an algorithm that computes the length of a $(1+\eps)$-approximate shortest path between $s$ and $t$
    in $O(1)$ rounds %in expectation and with high probability, 
    using $\Theta(\cS)$ space per machine and $O(n/\cS)$ machines where $\cS = n^{\delta}$ for any constant $\delta > 0$.
\end{corollary}

\subsection{Shortest cycle}
An application of $\eps$-emulators also allows us to find a $(1+\eps)$-approximate shortest cycle in an embedded planar graph.
Doing so is fairly simple.
If we partition the graph geometrically, either the shortest cycle lies completely in one piece of the partition or crosses a boundary vertex of some partition. Thus either we can find the shortest cycle on recursion, or we can compute an $\eps$-emulator on the boundary vertices and apply recursion. 

%Combined with our construction of the dual graph in \Cref{sec:dual}, this will give us an algorithm for global min-cut in an embedded planar graph.
\thmshortestcycle*
\begin{proof}
    We give a recursive algorithm.
    
    \paragraph{Base case.} 
    If $m\le \cS$ we can compute a $(1+\eps)$ approximate shortest cycle on one machine. The fastest algorithm for doing so takes $O_\eps(n)$ using $\eps$-emulators as sketched in \cite{ChangKT22}.
    
    \paragraph{Recursive case.}
    Let $B$ be a large bounding box that contains all of $G$.
We construct a $(r^{2/3}, 1/r)$-cutting-division $\Gamma$ for $r=\cS^\alpha$ for a sufficiently small $\alpha>0$ that splits $(G, \emptyset, B)$ into $k=O(r^{2/3})$ polygonal embedded graphs $(G_i, \nabla(G_i), P_i)$ for $i=1,...,k$. 

We can apply recursion to each $G_i$ independently to get the shortest cycle within $G_i$, discarding the outer boundary $P_i$. Let the length of the shortest cycle returned by this part be $\ell_1$.

Let $0 < \eps' < \eps$ be a parameter we will choose later.
We can also compute an $\eps'$-emulator $(H_i, \nabla(H_i), P_i)$ for each $(G_i, \nabla(G_i), P_i)$ using \Cref{lem:mpc_emulator} glue them together with the Gluing Lemma of \Cref{lem:gluing_lemma}.
This gives a graph $(H, \emptyset, B)$ with $\OO(m/r^{1/3})$ edges. We now apply recursion to $H$ to find a $(1+\eps')$-approximate shortest cycle of $H$. Call the length of the shortest cycle returned be $\ell_2$. Since $H$ approximates distances between vertices up to a distortion of $(1+\eps')$, $\ell_2$ is a $(1+\eps')^2 = (1+O(\eps'))$-approximate length of some (not necessarily simple or shortest) cycle in $G$. Choose $\eps' = \eps/C$ for a large enough constant $C$ so that $\ell_2$ is actually a $(1+\eps)$-approximation for the cycle.

We claim the length of the shortest cycle is $\min(\ell_1, \ell_2)$. 
Now we consider two cases of where the shortest cycle can be. Either it is completely contained in some graph $G_i$, in which case $\ell_1$ will be the shortest cycle.
Otherwise, the shortest cycle must cross the boundary of $\Gamma$.
The vertex where it crosses $\Gamma$ is a vertex of $\nabla(G_i)$, and hence, a vertex of $H$, so $\ell_2$ will be the length of a shortest cycle.

Note that at each level of the recursion, the $\eps$ we choose decreases by a factor of $1/C$. In the bottom level of the recursion, we are computing an $\eps'$-emulator with $\eps' = \eps/2^{O(1/\delta)}$. 
This is fine as $\delta$ is a constant.

It remains to analyze the round complexity of this algorithm. Let $T_{\circ}(m)$ denote the round complexity of finding the shortest cycle in an $m$ edge graph. We obtain the following recurrence.
\[ T_\circ(m) = T_\circ\left(\frac{m}{r^{1/3}}\right) + T_{emu}\left(\frac{m}{r^{1/3}}\right) + T_\circ\left(\OO\left(\frac{m}{r^{1/3}}\right)\right) + O(1) 
\]
As $T_{emu}(m) = O(1)$, this recurrence solves to $T_\circ (m) = O(1)$ for $r =\Theta(m^{\alpha \delta})$.
\end{proof}

\subsection{Single source shortest path}
With some additional ideas, it is actually possible to get all shortest paths from a source vertex $s \in V(G)$.
From \Cref{lem:mpc_emulator} we saw that it was possible to construct \emph{inside emulators} for each polygonal region, a graph that approximates all shortest paths inside the polygonal region.
The key idea is that we can actually construct \emph{outside emulators} for each polygonal region of the cutting-division after computing the inside ones.
For one polygonal region $P_i$, the outside consists of $O(1)$ connected regions that we can construct an $\eps$-emulator for from the other inside emulators that are outside $P_i$.
This idea is generally useful for many shortest path type problems. This allows us to prove the following theorem:

\thmsssp*

\begin{proof}
    We will prove the theorem by a recursive algorithm. Without loss of generality, we may assume $0< \eps < 1$.

\paragraph{Base Case.} If $m \le \cS$, we can use any sequential algorithm for computing approximate single source shortest paths. The fastest algorithm runs in $O(m)$ time using $\eps$-emulators (in fact, it solves a more general problem of multiple-source shortest paths)~\cite{ChangKT22}.

\paragraph{Recursive case.} 
We cannot use our standard choice of $s=r^{2/3}$ for our cutting-divisions, as we will get extra polylogarithmic factors from emulator constructions (for reasons we will shortly see).
We instead choose a slightly larger value of $s=r^{4/5}$ (though we could have chosen any exponent larger than $2/3$).

Let $B$ be a sufficiently large bounding box that contains all of $G$ with a small hole at $s$.
We construct a $(r^{4/5}, 1/r)$-cutting-division $\Gamma$ for $r= \cS^\alpha$ for a sufficiently small $\alpha>0$ that splits $(G, $\{s\}$, B)$ into $k=O(r^{1/5})$ polygonal embedded graphs $(G_i, \nabla(G_i), P_i)$ for $i=1,...,k$. Without loss of generality, let $G_1$ contain $s$ and that $s\in \nabla(G_1)$. Note that $s$ is on the boundary of some polygonal region because it was on the boundary of $B$.
In parallel, we apply \Cref{lem:mpc_emulator} on each polygonal embedded graph to get the inside emulators $(H_i, \nabla(G_i), Q_i)$ where each $(H_i, \nabla(G_i))$ are $\eps'$-emulators for $(G_i, \nabla(G_i))$ for some parameter $0 < \eps' < \eps$ we will choose later.

Next, we will show how to construct the outside emulators. 
Consider the region defined by $B\setminus P_i$, this decomposes into $\ell$ connected polygonal regions $R_1,..., R_\ell$.
Each $R_j$ for $j=1,...,\ell$ has at most $2$ holes, since $B$ had $2$ holes, one outside hole and one for $s$. 
Let $G_{R_j}$ denote the $R_j$ induced subgraph of $G$.
For all $R_j$, we will compute a distance emulator for $(G_{R_j}, \nabla(G_{R_j}), R_j)$ from the $\eps'$-emulators $(H_{i'}, \nabla(G_{i'}), Q_{i'})$ for $i' \neq i$ that we have already computed.
To do so, we observe that $R_j$ is the union of some collection of $P_{i'}$ for $i'\in I$, so we can view our cutting-division $\Gamma$ restricted on $R_{j}$ that we denote by $\Gamma_{R_j}$.
Thus, we can apply the Gluing Lemma, and apply \Cref{lem:mpc_emulator} on the glued graph to get a graph $(H_{R_j}, \nabla(G_{R_j}), Q_{R_j})$ which is an $\eps'$-emulator for $(G_{R_j}, \nabla(G_{R_j}, R_j)$. 
This can be done for all regions $R_j$ in parallel since the total number of edges in $H_{i'}$ for $i'\in I$ is $\OO(m/r^{2/5})$.
In fact we can do this for all $i$ in parallel since the total memory used by all $k = O(r^{1/5})$ graphs is only $k\cdot \OO(m/r^{2/5}) = \OO(m/r^{1/5})$. 

Now we can compute a redrawing of $G_i$ as $(G_i', \nabla(G_i), Q_i)$ by \Cref{lem:mpc_redrawing}.
We can apply the Gluing Lemma again on the polygonal embeddings $(H_{R_j}, \nabla(G_{R_j}), Q_{R_j})$ for $j=1,...,\ell$ along with $(G_i', \nabla(G_i), Q_i)$ to get a graph $(G_i'', \{s\}, B)$.
Note that $G_i''$ approximates all distances between pairs of vertices in $G$ by $(1+O(\eps'))$, and the number of edges in $G_i''$ is:
\begin{align*} |E(G_i'')| &= O(|E(G_i)|) + \OO\left(\sum_{j=1}^\ell |\nabla(G_{R_j})|\right) \\
&\le O(|E(G_i)|) + \OO\left(\sum_{i=1}^k |\nabla(G_i)|\right)  \\
&= O\left(\frac{m}{r^{1/5}}\right) + \OO\left(\frac{m}{r^{2/5}}\right) = O\left(\frac{m}{r^{1/5}}\right) 
\end{align*}
Hence, we can use recursion again to get all distances from $s$ in $(G_{i}'', \{s\}, B)$ up to $(1+\eps')$ distortion, which approximates the distances in $G$ with $(1+O(\eps'))$ distortion.
Note that we can recurse on all subproblems at once since the total size of all the subproblems is $k\cdot O(m/r^{1/5}) = O(m)$. 
Choose $\eps' = \eps/C$ for a sufficiently large constant $C$. This gives  $(1+\eps)$-approximate shortest path distances for all $P \in Q$.

Now to analyze the round complexity, let $T_{sssp}(m)$ denote the number of rounds to solve the SSSP problem with $m$ edges. Then the runtime follows the recursion:
\begin{align*}
T_{sssp}(m) 
&=  T_{emu}(m/r^{1/5}) + T_{emu}(\OO(m/r^{2/5})) + T_{draw}(m/r^{2/5}) + T_{sssp}(m/r^{1/5}) + O(1)
\end{align*}
As $r = \Theta(m^{\alpha \delta})$, $T_{emu}(m) = O(1)$ and $T_{draw} = O(1)$, this solves to $T_{sssp}=O(1)$.
\end{proof}

\subsection{All-pairs shortest paths}
A modification of our algorithm for computing approximate single source shortest path allows us to compute approximate all-pairs shortest path (APSP) in $O(1)$ rounds, at the expense of using $O(n^2)$ total memory for a $n$ vertex planar graph $G$. Note that this $O(n^2)$ total memory is much larger than the size of the graph which has $O(n)$ edges, but is necessary as we need $O(n^2)$ memory to even store all the pairwise distances.

In our algorithm for SSSP, we performed recursion on every component to compute shortest paths. The difference for APSP is that we will recurse once for each \emph{pair of components}.

\thmapsp*
\begin{proof}
We will prove the theorem by a recursive algorithm. We may assume $0< \eps < 1$. Let $m = |E(G)|$ and $n=|V(G)|$.

\paragraph{Base case.} If $m \le \cS$, we can use any sequential algorithm for computing approximate all-pairs shortest paths. This can easily be done in $O(n^2)$ time by running the SSSP algorithm in $O(n)$ time using $\eps$-emulators \cite{ChangKT22}.

\paragraph{Recursive case.} 
Let $B$ be a sufficiently large bounding box that contains all of $G$.
We construct a $(r^{4/5}, 1/r)$-cutting-division $\Gamma$ for $r= \cS^\alpha$ for a sufficiently small $\alpha>0$ that splits $(G, \emptyset, B)$ into $k=O(r^{1/5})$ polygonal embedded graphs $(G_i, \nabla(G_i), P_i)$ for $i=1,...,k$.
In parallel, we apply \Cref{lem:mpc_emulator} on each polygonal embedded graph to get the inside emulators $(H_i, \nabla(G_i), Q_i)$ where each $(H_i, \nabla(G_i))$ are $\eps'$-emulators for $(G_i, \nabla(G_i))$ for some parameter $0 < \eps' < \eps$ we will choose later. Note that this step only uses $O(m)$ total memory.

Next, we will construct the outside emulators for every pair of $(G_i, \nabla(G_i), P_i)$ and $(G_j, \nabla(G_j), P_j)$.
Consider the region defined by $B\setminus P_i\setminus P_j$, this decomposes into $\ell$ connected polygonal regions $R_1,..., R_\ell$.
Each $R_t$ for $t=1,...,\ell$ has at most $3$ holes, since $B$ had $3$ holes, one outside hole and at most two for $P_i$ and $P_j$.
For all $R_t$, we will compute a distance emulator for $(G_{R_t}, \nabla(G_{R_t}), R_t)$ from the $\eps'$-emulators $(H_{i'}, \nabla(G_{i'}), Q_{i'})$ for $i' \neq i, i'\neq j$ that we have already computed.
To do so, we observe that $R_t$ is the union of some collection of $P_{i'}$ for $i'\in I$, so we can view our cutting-division $\Gamma$ restricted on $R_{t}$ that we denote by $\Gamma_{R_t}$.
Thus, we can apply the Gluing Lemma, and apply \Cref{lem:mpc_emulator} on the glued graph to get a graph $(H_{R_t}, \nabla(G_{R_t}), Q_{R_t})$ which is an $\eps'$-emulator for $(G_{R_t}, \nabla(G_{R_t}, R_t)$. 
This can be done for all regions $R_t$ in parallel since the total number of edges in $H_{i'}$ for $i'\in I$ is $\OO(m/r^{2/5})$. This can also be done for all $O(r^{2/5})$ pairs $i$ and $j$ as this only uses $\OO(m)$ total memory.

Now we can compute a redrawing of $G_i$ and $G_j$ and glue it together with each of the $(H_{R_t}, \nabla(G_{R_t}), Q_{R_t})$, 
with the emulator of the rest of the graph as $(H_{ij}, \emptyset, B)$.
Note that $G_{ij}$ approximates all distances between pairs of vertices in $G$ by $(1+O(\eps'))$, and the number of edges in $G_{ij}$ is:
\begin{align*} |E(G_{ij})| = O(|E(G_i)| + |E(G_j)| + \OO\left(\sum_{j=1}^\ell |\nabla(G_{R_j})|\right) 
= O\left(\frac{m}{r^{1/5}}\right) + \OO\left(\frac{m}{r^{2/5}}\right) 
= O\left(\frac{m}{r^{1/5}}\right) 
\end{align*}
Hence, we can use recursion again to get all distances from $s$ in $(G_{ij}, \emptyset, B)$ up to $(1+\eps')$ distortion, which approximates the distances in $G$ with $(1+O(\eps'))$ distortion.

Let us analyze the total space used by our algorithm. Let $M_{apsp}(m)$ denote the memory usage of our algorithm for a $m$ edge graph $G$ with no isolated vertices. We get the following recurrence.
\begin{align*} M_{apsp}(m) &= \sum_{i=1}^k\sum_{j=i+1}^k M\left(|E(G_{ij})|\right) + \OO(m)
= O(r^{2/5}) \cdot M_{apsp}\left(\frac{m}{r^{1/5}}\right) + O(m)
\end{align*}
Along with the base case that $M_{apsp}(\cS) = O(\cS^2)$ this solves to $M_{apsp} = O(m^2) = O(n^2)$.

Now to analyze the round complexity, let $T_{apsp}(m)$ denote the number of rounds to solve the APSP problem with $m$ edges. Then the runtime follows the same recursion as for SSSP:
\begin{align*}
T_{apsp}(m) &=  T_{emu}(m/r^{1/5}) +  T_{emu}(\OO(m/r^{2/5})) + T_{draw}(m/r^{1/5}) + T_{apsp}(m/r^{1/5}) + O(1)  
\end{align*}
As $r = \Theta(m^{\alpha \delta})$, $T_{emu}(m) = O(1)$ and $T_{draw} = O(1)$, this solves to $T_{apsp}(m)=O(1)$.
\end{proof}

\subsection{Drawing the primal-dual overlay graph} \label{sec:dual}
\begin{figure}
    \centering
    \includegraphics[height=0.2\textheight, page=1]{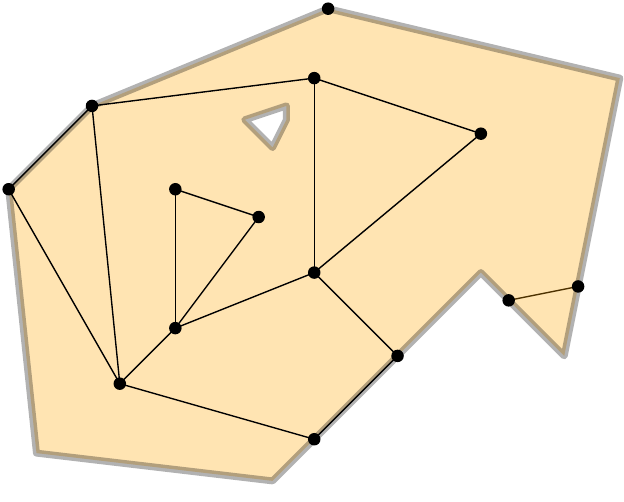}
    \includegraphics[height=0.2\textheight, page=2]{primaldualoverlay}
    \includegraphics[height=0.2\textheight, page=3]{primaldualoverlay}
   \caption{
   \textbf{(Left)} A polygonal embedding $(G, \nabla(G), P)$.
   \textbf{(Middle)} A primal-dual overlay graph of one component of the graph. The green square vertices result from subdividing the edges. The red edges form a subdivision of the dual graph.
   \textbf{(Right)} The polygonal primal-dual overlay graph $(G^{\PD}, \nabla(G) + \nabla(G^*), P)$ with the parts of the polygon parallel to edges slightly pushed outward for visual clarity.
   }
   
    \label{fig:primal-dual_overlay}
\end{figure}

\paragraph{Defining the primal-dual overlay graph.}
The \emph{planar dual} of an embedded planar graph $G$ is the graph $G^*$ obtained by creating a vertex for every face of $G$ we call a \emph{face vertex} and connecting faces that share an edge.
Each edge of the primal planar graph $G$ corresponds to exactly one edge of $G^*$. Furthermore, each face of $G^*$ corresponds to a vertex of the primal graph.
We will also define the \emph{primal-dual overlay graph} to be a graph $G^{PD}$ that combines both the primal and the dual graph. 
It can be constructed from an embedded planar graph in this way: 
Subdivide all the edges of $G$ creating a vertex $v_e$ for every edge $e\in E(G)$. We call $v_e$ an edge vertex. Place a vertex $f$ in each face of $G$ and connect $f$ to each subdivided vertex $v_e$ that the face shares an edge $e$ with.
Note that this graph is a planar graph that contains both $G$ after subdividing every edge and $G^*$ after subdividing every edge as subgraphs. 
Furthermore, if the graph $G$ had $n=|V(G)|$ vertices, since the number of faces and edges of $G$ is also $O(n)$, observe that $|V(G^{PD})| = O(n)$. See \Cref{fig:primal-dual_overlay} for an illustration.

We also define the \emph{primal-dual overlay of a polygonal embedding $(G, \nabla(G), P)$}. We construct this graph as follows:
Consider each hole of $P$, and connect all adjacent vertices of $\nabla(G)$ with an edge even if one already exists and is parallel to the edge (consider expanding the boundary slightly outward so the edge is no longer parallel) making a planar multigraph $G'$. If there are no vertices on the boundary of the hole, add a phantom vertex on the hole which has a self edge. Construct the primal-dual overlay graph of $G'$ and delete all vertices corresponding to holes of $H$ and their incident edges.
The vertices created by subdividing the edges of $P$ we consider as $\nabla(G^*)$, the boundary of the dual graph. We call the graph without the (subdivided) edges of $P$ and phantom vertices, $G^{PD}$. 
The entire polygonal embedding is denoted by $(G^{PD}, \nabla(G) + \nabla(G^*), P)$. See \Cref{fig:primal-dual_overlay} for an illustration of this.

The motivation for this construction is if we had a graph $G$ and partition of the plane $\Pi$ that induced polygonal embeddings $(G_i, \nabla(G_i), P_i)$ for $i=1,...,k$, then the union of the primal-dual overlay of the polygonal embeddings $(G^{PD}_i, \nabla(G_i) + \nabla(G_i^*), P_i)$ contains $G^{PD}$ as a minor provided we identify $\nabla(G_i^*)$ along the boundaries between adjacent regions.

\paragraph{Computing the primal-dual overlay graph.}
We can in fact modify our method of redrawing a graph in \Cref{lem:mpc_redrawing} to compute a graph $H$ that contains the primal-dual overlay graph $G^\PD$ as a minor.
We note some properties of $H$.
Our technique of computing a cutting-division may cut faces (corresponding to a vertex of $G^*$) into parts.
We will create a vertex on the boundary of the cutting-division for the face that we are cutting. 
When we glue the faces back together we will connect the faces together by an edge. This has the effect of doing a more general kind of vertex split of a facial vertex; it is more general because a single cutting-division may cut up a face into multiple parts such that all the vertices are connected in a planar manner.
A face of $G$ (i.e., a vertex of $G^*$) will thus correspond to a connected set of face vertices in $H$.
We note that the edges of $H$ can be decomposed into two graphs $H'$ and $H^*$, where $H'$ is the induced subgraph of $H$ with the edge vertices and the old vertices of $G$ and thus has $G$ as a minor and $H^*$ is the induced subgraph of $H$ with the edge vertices and face vertices and thus has $G^*$ as a minor.
We refer to $H'$ as the \emph{primal subgraph} of $H$, and $H^*$ as the \emph{dual subgraph} of $H$.

The high-level idea of how we will construct the primal-dual overlay graph is to find a cutting-division of the graph, and recursively draw the primal-dual overlay graph of the polygonal subdivision.

\begin{lemma}[MPC drawing of the primal-dual overlay graph] \label{lem:mpc_overlay}
    Given an embedded planar graph $G$ with $n$ vertices and $m$ edges,
    we can compute an embedded planar graph $H$ with $O(m)$ edges that contains both the primal graph $G$ and the dual graph $G^*$ as a minor 
    in $O(1)$ rounds 
    %in expectation and with high probability, 
    using $\Theta(\cS)$ space per machine and $O(n/\cS)$ machines where $\cS = n^{\delta}$ for any constant $\delta > 0$.
\end{lemma}

For a graph $G$, 
let $B$ be a sufficiently large bounding box that contains all of a graph $G$.
We will apply the below claim to $(G, \emptyset, B)$, which immediately proves our theorem.
\begin{claim} \label{clm:mpc_overlay}
Let $r = \cS^{\alpha}$ for a sufficiently small $\alpha>0$.
There exists an algorithm that 
given an $O(1)$-holed polygonal embedded graph $(G, \nabla (G) + \nabla(G^*), P)$ (i.e., the polygonal embedding with isolated vertices of the primal-dual overlay graph of the polygonal embedding on the boundary)
with $n$ vertices and $m$ edges 
where $|\partial(P)|= O(r^{1/3})$,
returns a polygonal embedded graph $(H, \nabla(G) + \nabla (G^*), Q)$
satisfying these properties:
\begin{itemize}
    \item $\nabla(G^*)$ is a set of vertices of $G^*$ to faces of $G$ lying on the boundary of $P$
    \item $V(H) \supseteq V(G) \cup  V(G^*)$
    \item $|H| = O(m)$ 
    \item $H$ contains the primal-dual overlay of the polygonal embedding $G^{PD}$ as a minor
    \item Each edge $e\in E(H)$ corresponds to either a canonical edge of $G$, a virtual edge from a vertex split or subdivision of $G$,
    the canonical dual edge in $G^*$, a virtual edge from a vertex split or subdivision of $G^*$ and stores this information.
    \item Each vertex of $H$ corresponds to either the vertex of $G$, a virtual vertex from a subdivision of an edge of $G$, a vertex of $G^*$ (i.e., a face of $G$), or is a virtual vertex from a vertex split of $G^*$ and stores this information.
\end{itemize}
The algorithm performs in $O(1)$ rounds 
%with high probability and in expectation
using $\Theta(\cS)$ space per machine and $O(n/\cS)$ machines where $\cS = n^{\delta}$ for any constant $\delta > 0$.
\end{claim}
\begin{proof}

We will use a recursive algorithm to draw the primal-dual overlay in a manner similar to how we redrew the graph in \Cref{lem:mpc_redrawing}.
\paragraph{Base case.} If $m\le \cS$, then we can compute the primal-dual overlay on one machine. 
We can explicitly compute $(H, \nabla(G) + \nabla(G^*), P)$ where $H$ is the primal-dual overlay graph of $G$.
%This 
We can draw in a triangular domain with the Redrawing Lemma (\Cref{lem:redrawing}).

\paragraph{Recursive case.}
We construct a $(r^{2/3}, 1/r)$-cutting-division $\Gamma$ of $(G, \nabla(G)+\nabla(G^*), P)$ with $r=\cS$ which induce embeddings $(G_i, \nabla(G_i), P_i)$. 
For each $i$, we can compute $\nabla(G_i^*)$, the vertices of $G^*_i$ we place along the boundary of $P$ explicitly and in parallel for each region.
Begin by adding all degree $3$ or higher vertices of $\Gamma$ to the boundaries of all regions (this is to maintain consistency across the boundaries of the regions).
We proceed by sorting $\nabla(G_i)$ around each boundary based on distance along each hole from an arbitrary base point. 
This can be done in $O(1)$ rounds as all machines have $\Gamma$.
Next, for every consecutive vertex $u$ and $v$ of $\nabla(G_i)$ on each boundary, we place a vertex $w\in \nabla(G^*_i)$ in the exact midpoint between $u$ and $v$ on the boundary.
Note that on the other side of the boundary, some other polygonal region $P_j$ will have put a vertex in the same spot.
This vertex $w$ corresponds to a face vertex of $G^*_i$.
If a boundary hole of $P_i$ has no vertices of $\nabla(G_i)$,  we can deterministically place a vertex $w\in \nabla(G^*_i)$ in a canonical location (say the bottom-left most vertex of the hole).

Now we can recurse on each $(G_i, \nabla(G_i)+\nabla(G^*_i), P_i)$, and glue together the returned polygonal embeddings. The analysis follows from \Cref{lem:mpc_redrawing}.
The runtime analysis is the same as well since for every component $|\nabla(G_i^*)| = O(|\nabla(G_i)| + |\Gamma|)$.
\end{proof}

\paragraph{Correcting the dual subgraph.} We state a few quirks of the dual subgraph $H^*$ resulting from our use of constructing $H$ while subdividing edges and faces of $G$. The vertices corresponding to a single face of $G$ will be connected but may have cycles. This happens because we may cut up large faces into many pieces. For long edges that we subdivide many times, we may connect multiple vertices of $H^*$ corresponding to the same face of $G$ to different subdivisions of the same edge. 

We will do some post-processing on $H^*$ to recover a more useful representation of the dual graph.
We begin by removing cycles among facial vertices that correspond to the same face of $G$ by computing a spanning forest for each dual face in $O(1)$ rounds using \Cref{thm:mst}. 
To handle face vertices corresponding to the same face connected to edge vertices corresponding to the same edge, we will remove duplicate edges by only keeping the edge from the face vertex to the edge corresponding to the canonical edge vertex. This can also be done in $O(1)$ rounds. 
Let $\oH^*$ be the graph we get from this procedure.
We will call $\oH^*$ the \emph{corrected dual subgraph} of $H$.

In particular, if we look at any cycle $C$ using edges of $\oH^*$, the cycle corresponds to an actual cycle in $G^*$ and thus a cut in $G$.
Furthermore, the vertices on one side of the cut are exactly the vertices of $H$ that correspond to vertices of $G$ that lie in the cycle.
We will use this property for computing max-flows in \Cref{sec:mincut}.

\subsection{Min-cut and max-flow} \label{sec:mincut}
Our dual graph construction, combined with an algorithm for $(1+\eps)$-approximate shortest cycle, gives an algorithm for global min-cut, as the min-cut of a planar graph is the shortest cycle in the dual.

\thmmincut*
\begin{proof}
The min-cut is the shortest cycle of the dual graph we compute using \Cref{lem:mpc_overlay} if we only look at the dual part of the primal-dual overlay and remove cycles corresponding to faces with \Cref{thm:mst}, then 
\Cref{thm:shortest_cycle} computes the approximate shortest cycle.
\end{proof}

Computing a max-flow between two terminals $s$ and $t$ in $G$ is more challenging, as this is equivalent to computing the minimum cut in the dual $G^*$ that separates the two \emph{faces} $s^*$ and $t^*$. Nonetheless, we can prove the following theorem by adapting ideas for computing inside and outside emulators.

\thmmaxflow*

To prove the max-flow theorem, we first compute an embedding of the primal-dual overlay graph $H$ using \Cref{lem:mpc_overlay} of the component of the graph that $s$ and $t$ are in (we can figure out if they are connected and get all edges and vertices of that component using variations of \Cref{thm:cc}).
Instead of storing the entire dual face of $s^*$ and $t^*$, which may have many edges, we will simply store the coordinates of the two vertices in the primal-dual overlay graph as a proxy for the face.
Let $p_1$ denote the location of $s$ and $p_2$ the location of $t$ in $H$. Use as input to the following lemma the corrected dual subgraph $\oH^*$, as well as $p_1$ and $p_2$ to prove \Cref{thm:maxflow}.
\begin{lemma}
    There exists an algorithm that 
    given as input an embedded planar graph $G$ with $n$ vertices and $m$ edges and two points $p_1$ and $p_2$ not lying in the same face of $G$,
    can find a $(1+\eps)$-approximate minimum cycle of $G^*$ 
    that separates $p_1$ and $p_2$
    in $O(1)$ rounds 
    %in expectation and with high probability, 
    using $\Theta(\cS)$ space per machine and $O(n/\cS)$ machines where $\cS = n^{\delta}$ for any constant $\delta > 0$.
\end{lemma}

\begin{proof} 
We assume that $p_1$ and $p_2$, and all coordinates of $G$ have different $x$ coordinates. If not, we can rotate all points by a random angle.
We prove this theorem by a recursive algorithm very similar to \Cref{thm:sssp}. 
The idea the shortest cycle separating $p_1$ and $p_2$ will either pass through the boundary of a cutting-division, or will be completely contained in a single polygonal region of the cutting-division, say the $i$th region (i.e., it can contain the polygonal region containing $p_1$ as a hole, and have $p_2$ outside. In either case, if we glue together a redrawing of each region with redrawn regions containing $p_1$ and $p_2$, treating them as holes so the topology of the graph does not change, then also compute an outside emulator of everything other than these three regions, and recurse on this other graph, we will find the shortest cycle.

\paragraph{Base case.} If $m\le \cS$ we can solve this problem on one machine in $O(n)$ time by the algorithm of Chang--Krauthgamer--Tan~\cite{ChangKT22}.

\paragraph{Recursive case.} 

Let $B$ be a sufficiently large bounding box that contains all of $G$ with punctures at $p_1$ and $p_2$.
We construct a $(r^{1/5}, 1/r)$-cutting-division $\Gamma$ for $r=\cS^\alpha$ with $\alpha >0$ sufficiently small on $(G,\emptyset, B)$ that splits the problem into polygonal embedded graphs $(G_1,\nabla(G_1), P_1),..., (G_k,\nabla(G_k), P_k)$ for $k=O(r^{1/5})$, and construct 
a redrawing $(H_i, \nabla(G_i), Q_i)$ and
an $\eps'$-emulator $(H_i', \nabla(G_i), Q_i)$ for each $i=1,...,k$ for $\eps' = \eps/C$ for a sufficiently large $C$.

We may assume that the cutting-division does not contain $p_1$ or $p_2$ as we assume that $p_1$ and $p_2$ have different $x$-coordinates.
Let $P_1$ be the polygonal region containing $p_1$, and let $P_j$ be the polygonal region containing $p_2$. Note that it is possible that $j=1$.
We will describe how we do this for one specific region $i>1$ and $i\neq j$, we construct an outside emulator for $B\setminus (P_1 \cup P_j \cup P_i)$, from the $\eps'$-emulators we already constructed. 
Let $(\bar{H}_i, \nabla(\bar{H}_i), B\setminus (P_1 \cup P_j \cup P_i))$ be the emulator.
Now, we can glue together that emulator with the redrawn $(H_1, \nabla(G_1), Q_1)$, $(\bar{H}_i, \nabla(G_i), Q_i)$, and $(H_j, \nabla(G_i), Q_j)$ to get a single graph $\bar{H}$ in a region $Q'$ with three holes: one outside boundary, and one hole for each of $p_1$ and $p_2$. Note that in $Q'$, the holes corresponding to $p_1$ and $p_2$ may become triangles in the redrawing, but have no boundary vertices. We can choose any vertex of the triangle $p_1'$ from the hole corresponding to $p_1$ and $p_2'$ from the hole corresponding to $p_2$. Now we can recurse on the graph $\bar{H}$ with $p_1'$ and $p_2'$. 
For the correctness of this algorithm, it is clear that a cycle that separates $p_1'$ and $p_2'$ in $\bar{H}$ corresponds to a cycle that separates $p_1$ and $p_2$ in $G$. 

$\bar{H}$ has $O(m/r^{1/5})$ edges, we need to repeat this construction for each $i$, so we have $O(r^{1/5})$ such subproblems to solve. We can solve all these problems in parallel.
The analysis of runtime is the same as for \Cref{thm:sssp}, but we will describe it here for completeness.
Let $T_{mincut}(m)$ denote the number of rounds to find the shortest $p_1$ and $p_2$ separating cycle in a graph with $m$ edges. Then the runtime follows the recursion:
\begin{align*}
T_{mincut}(m) 
&=  T_{emu}(m/r^{1/5}) + T_{emu}(\OO(m/r^{2/5})) + T_{draw}(m/r^{2/5}) + T_{mincut}(m/r^{1/5}) + O(1)
\end{align*}
As $r= \Theta(m^{\alpha\delta})$, $T_{emu}(m) = O(1)$ and $T_{draw} = O(1)$, so this solves to $T_{mincut}=O(1)$.
\end{proof}

\section{Edit distance}\label{sec:edit_distance}

%In the longest common subsequence (LCS) problem, we are given two strings $s$ and $t$ and we wish to find the longest string that both $s$ and $t$ contain as (not necessarily consecutive) subsequences.
In the edit distance problem, we are given two strings $s$ and $t$, and we wish to determine the minimum number of operations to transform $s$ to $t$. In an operation we may insert, delete, or change a character at any position of $s$. 
We also consider a weighted variation of the problem where we may have varying costs for inserting, deleting, or replacing certain characters in the strings $s$ and $t$. 

Formally we define a cost function $w: (\Sigma \cup \{\emptyset\}) \times (\Sigma \cup \{\emptyset\}) \to \R^+$ where $\Sigma$ is the alphabet, and $w(a, \emptyset)$ is the cost of inserting a character $a$ and $w(\emptyset, b)$ is the cost of deleting a character $b$, and $w(a,b)$ is the cost for replacing $a$ with $b$ for $a,b\in \Sigma$. 

A weight function $w$ is \emph{quasimetric} if 
it satisfies the triangle inequality 
$w(a,c) \le w(a, b) + w(b, c)$ for all $a,b,c\in \Sigma \cup\{\emptyset\}$. %\yijun{I think we just need to consider the case where $c = \emptyset$ and $a, b \in \Sigma$, but I guess this is equivalent to the more general case that you stated?}
It is reasonable to assume we always deal with quasimetric weight functions,
since we can always do edit distance computations
on a weight function $\ow$ where $\ow(a,c)$ is the cost of the cheapest way to transform $a$ to $c$.
We will focus on the setting where the alphabet size is a constant, so the preprocessing costs constant time and can be done locally.
We say that a weight function $w$ is \emph{symmetric} if $w(a,b) = w(b,a)$ for all $a,b\in \Sigma \cup \{\emptyset\}$. In particular, this means that the cost of inserting and deleting a character is the same.
In the classic edit distance problem, all weights are $1$ and the weights are symmetric and quasimetric.

The textbook dynamic programming algorithm for edit distance creates a table $A$ with the following recurrence if we let $s_i$ denote the $i$th character of string $s$ and $t_j$ denote the $j$th character of string $j$.
\[
A[i][j] =
\begin{cases}
-\infty \qquad \text{ if $i<0$ or $j<0$} \\
0 \;\;\;\;\;\qquad \text{ if $i=j=0$} \\
%A[i-1][j-1] & \text{ if $s_i=t_j$} \\
\min\{w(s_i,\emptyset) + A[i-1][j], w( \emptyset, t_j) +A[i][j-1], w(s_i, t_j) + A[i-1][j-1]\}
\end{cases}
\]
\newcommand{\Gwst}{G^w_{s,t}}
\newcommand{\oGwst}{\overline{G}^w_{s,t}}
\newcommand{\oGwstprime}{\overline{G}^w_{s',t'}}

Another interpretation of this dynamic program is that it
defines a directed graph $\Gwst$
with vertices $(i,j) \in V$ for $i = 1, \dots, |s|$ and $j = 1, \dots, |t|$.
For a vertex $(i,j)$ there are incoming edges:
\begin{itemize}
\item from $(i-1, j)$ with weight $w(s_i, \emptyset)$ for corresponding to a deletion of $s_i$,
\item from $(i, j-1)$ with weight $w(\emptyset, t_j)$ corresponding to an insertion of $t_j$,
\item from $(i-1, j-1)$ with weight $w(s_i, t_j)$ corresponding to a replacement of $s_i$ with $t_j$.
\end{itemize}
The table $A[i][j]$ stores the shortest path in this graph from $(0,0)$ to $(i,j)$.
We remark that $G^w_{s,t}$ is a planar graph 
with a straight-line embedding if we place vertices at the corresponding $x$ and $y$ coordinates.

We observe that we can actually view this as an \emph{undirected} shortest path problem when the weights function is quasimetric and symmetric, and work with the graph $\oGwst$ that has the same vertex set and edge set as $\Gwst$, except the edges are undirected.

\begin{observation} \label{obs:undirecting_g}
For symmetric and quasimetric weight functions, the shortest path between $(0,0)$ and $(i, j)$ in $\Gwst$ is equal to the shortest path between those vertices in $\oGwst$.
\end{observation}
\begin{proof}
It is sufficient to show that there exists a shortest path from $(0,0)$ to $(i,j)$ in $\oGwst$ that is non-decreasing in both dimensions. We select $P$ as a shortest path from $(0,0)$ to $(i,j)$ in $\oGwst$ such that its hop-length is minimized. 
If $P$ is non-decreasing in both dimensions, then we are done.
For the rest of the proof, suppose $P$ is not non-decreasing in the $x$-dimension or in the $y$-dimension. 

We select $b=(x_b, y_b)$ as the \emph{first} vertex on $P$ such that the edge leading to $b$ in $P$ is a decreasing move in at least one dimension. Let $b'=(x_{b'}, y_{b'})$ be the vertex on $P$ right before $b$. Then  $(x_{b'} = x_{b} + 1) \vee (y_{b'} = y_{b} + 1)$ is true.

\paragraph{Case 1: $b' \rightarrow b$ is diagonal.}
We first consider the case where $b' = (x_b+1, y_b+1)$. That is, the edge between $b'$ and $b$ is diagonal.  Let $b''$ be the vertex on $P$ right before $b'$. Our choice of $b$ implies that $b''=(x_{b}, y_{b}+1)$ or $b''=(x_{b}+1, y_{b})$. Now we may shorten the hop-length of $P$ by replacing $b'' \rightarrow b' \rightarrow b$ with $b'' \rightarrow b$. We claim that the weight of the new path cannot be larger than the weight of the old path. 

We only prove this claim for the case where $b''=(x_{b}, y_{b}+1)$, as the other case is similar. The weight of $b'' \rightarrow b' \rightarrow b$ equals $w(s_{x_b+1}, \emptyset) + w(s_{x_b+1}, t_{y_b+1})$, and the weight of $b'' \rightarrow b$ equals $w(\emptyset, t_{y_b+1})$. We have
\[
w(s_{x_b+1}, \emptyset) + w(s_{x_b+1}, t_{y_b+1}) \geq w(t_{y_b+1}, \emptyset) = w(\emptyset, t_{y_b+1}),
\]
where the inequality is due to the fact that $w$ is quasimetric and the equality is due to the fact that $w$ is symmetric.

\paragraph{Case 2: $b' \rightarrow b$ is horizontal or vertical.}
Next, we consider the case where $b' = (x_b + 1, y_b)$ or $b' = (x_b, y_b + 1)$. That is, the edge between $b'$ and $b$ is horizontal or vertical. For the rest of the proof, we only focus on the case where $b' = (x_b, y_b + 1)$, since the other case is similar. 

We select $a=(x_a, y_a)$ to be the \emph{last} vertex on $P$ before $b$ such that $y_a = y_b$. Let $a'=(x_{a'}, y_{a'})$ be the vertex on $P$ right after $a$. Our choice of $a$ implies that we must have $y_{a'} = y_a + 1$, so either $a' = (x_a, y_a + 1)$ or $a' = (x_a+1, y_a + 1)$. For the rest of the discussion, we write $k = y_a = y_b$ for notational simplicity.

\paragraph{Case 2.1: $a \rightarrow a'$ is vertical.} Consider the case where  $a' = (x_a, y_a + 1)$. Recall that $y_a = y_b = k$, so $y_a+1 = y_{a'} = y_{b'} = y_b+1 = k+1$, meaning that the subpath of $P$ from $a'$ to $b'$ is a horizontal line segment from $(x_a,k+1)$ to $(x_b,k+1)$, so the subpath of $P$ from $a$ to $b$ is
\[a=(x_a, k) \rightarrow (x_a, k+1) \rightarrow (x_a +1, k+1) \rightarrow (x_a +2, k+1) \rightarrow \cdots \rightarrow (x_b , k+1) \rightarrow (x_b , k)=b.\]
We can shorten $P$ by replacing this subpath with the path
\[a=(x_a, k)  \rightarrow (x_a +1, k) \rightarrow (x_a +2, k) \rightarrow \cdots  \rightarrow (x_b, k)=b.\]
The weight of the new path is at most the weight of the old path because
 the weight of $(i-1, k) \rightarrow (i, k)$ is identical to the weight of $(i-1, k+1) \rightarrow (i, k+1)$ for all $i$, as they are both $w(s_i, \emptyset)$. This contradicts our choice of $P$, as the hop-length of the new path is smaller than the hop-length of the old path.

\paragraph{Case 2.2: $a \rightarrow a'$ is diagonal.} Consider the case where  $a' = (x_a+1, y_a + 1)$. Similarly, the subpath of $P$ from $a'$ to $b'$ is a horizontal line segment from $(x_a+1,k+1)$ to $(x_b,k+1)$, so the subpath of $P$ from $a$ to $b$ is
\[a=(x_a, k) \rightarrow (x_a +1, k+1) \rightarrow (x_a +2, k+1) \rightarrow \cdots \rightarrow (x_b , k+1) \rightarrow (x_b , k)=b.\]
Similarly, we can shorten $P$ by replacing this subpath with the path
\[a=(x_a, k)  \rightarrow (x_a +1, k) \rightarrow (x_a +2, k) \rightarrow \cdots  \rightarrow (x_b, k)=b.\]
We claim that the weight of the new path is at most the weight of the old path. 
As discussed earlier, we already know that the weight of $(i-1, k) \rightarrow (i, k)$ is identical to the weight of $(i-1, k+1) \rightarrow (i, k+1)$ for all $i$, so we just need to prove that the weight of $(x_a, k)  \rightarrow (x_a +1, k)$, which equals $w(s_{x_a +1}, \emptyset)$, is at most the sum of the weight of $(x_a, k) \rightarrow (x_a +1, k+1)$ and the weight of $(x_b , k+1) \rightarrow (x_b , k)$, which equals $w(s_{x_a +1}, t_{k+1}) + w(\emptyset,t_{k+1})$. Indeed,
\[w(s_{x_a +1}, \emptyset) \leq w(s_{x_a +1}, t_{k+1}) + w(t_{k+1}, \emptyset) = w(s_{x_a +1}, t_{k+1}) + w(\emptyset,t_{k+1}),\]
where the inequality follows from the assumption that $w$ is quasimetric and the equality follows from the assumption that $w$ is symmetric.
This contradicts our choice of $P$, as the hop-length of the new path is smaller than the hop-length of the old path.

Since a contradiction is obtained in all cases, $P$ must be non-decreasing in both dimensions.
\end{proof}

As a warmup, we will show an MPC algorithm using $O(n^2)$ total memory that is a corollary of our result on approximate SSSP.
\begin{corollary}
There exists an algorithm that 
given as input  two strings $s$ and $t$ of length $n$ and a symmetric quasimetric weight function $w$, 
computes a $(1+\eps)$-approximate edit distance between $s$ and $t$
in $O(1)$ rounds using $O(S)$ space per machine
and $O(n^2/S)$ machines where $S = n^{\delta}$ for any constant $\delta > 0$.
\end{corollary}
\begin{proof}
We begin by partitioning $s$ and $t$ into contiguous substrings of length $\ell = \sqrt{\cS}$.
For simplicity, we assume $\sqrt{\cS}$ is an integer and $m = (n-1)/\ell$ is an integer.
Let $s^{(i)}$ denote the string $s_{(i-1)\ell+1}s_{(i-1)\ell+2}\cdots s_{i\ell+1}$ for $i\in [m]$, and similarly for $t^{(i)}$.

For each $i, j \in [m]$, we distribute $s^{(i)}$ and $t^{(j)}$ to a different machine along with the 
weight function $w$. Observe that we can do this as we have $O(m^2) = O(n^2/\cS)$ machines.
This allows us to construct $\oGwst$ as every machine constructs an $O(\cS)$ sized \emph{block} of $\oGwst$.
Formally, we say the $(i,j)$ block of $\oGwst$ is the set of edges
between vertices $(a, b)$ for $a\in [(i-1)\ell+1, i\ell + 1]$ and $b=[(j-1)\ell+1, j\ell + 1]$ of $\oGwst$.
Putting all the edges together (and removing duplicates), we have computed all edges of $\oGwst$.

Now we can use our result on approximate SSSP in planar graphs \Cref{thm:sssp} 
to compute a $(1+\eps)$-approximate shortest path from $(0,0)$ to $(n,n)$.
By \Cref{obs:undirecting_g}, this is a $(1+\eps)$-approximate weighted edit distance.
\end{proof}
We remark that as the entire graph is a grid graph, the entire algorithm can be significantly simplified (e.g. cutting-divisions can be blocks of the grid).

We are able to match the result of Hajiaghayi, Seddighin, and Sun~\cite{HajiaghayiSS19}
where they use only $\OO(n^2/\cS^2)$ machines.
This takes a little extra effort, as with fewer machines, we need 
each machine to take a longer $\cS$ sized substring of $s$ and $t$.
Nonetheless, we show that we can construct a sparser version of $\oGwst$ 
that approximately preserves distances with this much memory.

\thmeditdist*

As before we can distribute pairs of contiguous substrings, one from each string, and the weight function $w$ to every machine. However, this time, we need to set the length $\ell = \cS$, so we have $m=(n-1)/\ell = O(n/\cS)$ contiguous substrings of each string.
This way we can still distribute $s^{(i)}$ and $t^{(j)}$ to a different machine for each $i,j\in [m]$ as now we only need $O(m^2) = O(n^2/\cS^2)$ machines.

Now we're faced with a dilemma. The machine given $s^{(i)}$ and $t^{(j)}$ has two length $\cS$ strings but we only have $O(\cS)$ total memory.
The $(i,j)$ block of the graph $\oGwst$ has $O(\cS)$ memory.
Fortunately, we can use $\eps$-emulators to construct a sparser graph of size $\OO(\cS)$ on the $O(\cS)$ boundary vertices 
of the block of the graph. If we were able to do this, we could run the deterministic version of \Cref{thm:sssp} which computes the single source shortest path with $O(\eta^{1+\alpha})$ work on an $\eta$ vertex graph for any constant $\alpha > 0$. As $\eta = \OO(n/\cS)$, choosing a constant $\alpha < \delta$ would solve the problem with $o(n^2)$ total work.

However, it is difficult to construct an $\eps$-emulator for the $(i,j)$ block
without being able to store the entire block to begin with!
To resolve this issue, we will use a simple recursive algorithm to do exactly this with divide and conquer. To do so, we will prove the following lemma from which \Cref{thm:edit_dist} follows.

\begin{lemma}
Fix a parameter $\eps$ with $0<\eps \le 1/2$.
Given two strings $s$ and $t$ of length $n$, there exists an algorithm that computes an $\eps$-emulator for $\oGwstprime$ with boundary vertices $(x,0)$, $(0, y)$, $(x, |s|)$, and $(|t|, y)$ for all $x,y \in [n]$ in $\OO(n^2)$ time and $\OO(n)$ memory.
\end{lemma}

\begin{proof}
We begin by splitting $s$ in half into $s^{(1)}$ and $s^{(2)}$, and $t$ in half into $t^{(1)}$ and $t^{(2)}$.
We will compute an emulator on
each of $s^{(i)}$ and $t^{(j)}$ for $i,j \in\{1,2\}$
of size $\OO(n/2)$ with $\OO(n/2)$ space by recursion, and glue them together at the boundary vertices
for a single graph $H$ with the techniques from \Cref{sec:gluing} (we can view our partition as a particularly simple cutting-division).
Then, we can reduce the size of $H$ using an $\eps'$-emulator for some $\eps'$ to be specified later to get a graph $H'$. 
$H'$ has size $\OO(n)$ and can be computed in $\OO(n)$ time and space by \Cref{thm:eps_emulator}.
We repeat this recursively until our subproblems have size $O(\sqrt{n})$, we can afford to directly compute an $\eps'$ emulator directly with $\OO(n)$ space,
so our recursion has depth $(\log n)/2$.
It remains to argue that the final graph we obtain is an $\eps$-emulator.
Observe that gluing together two $\gamma$-emulators results in an $\gamma$ emulator.
Taking an $\gamma$-emulator of an $\gamma'$-emulator gives a graph with distortion at most $(1+\gamma)\cdot (1+\gamma')$.
If we choose $\eps'$ sufficiently small such that $ \eps = O(\eps'\log n)$, we can bound the overall approximation factor as follows:
\[ \prod_{i=1}^{(\log n)/2} (1+\eps') 
%\le \prod_{i=1}^{(\log n)/2} e^{\eps'} = e^{(\eps'\log n)/2} 
\le 1+ O(\eps' \log n) \le 1+\eps \]
Observe that even when we choose an $\eps' = \Theta(1/\log n)$, an $\eps'$-emulator of an $n$ vertex graph still has $\OO(n)$ size and can be computed in $\OO(n)$ time and space by \Cref{thm:eps_emulator}.

Since we do computation on each recursive subproblem in sequence, we can reuse the space.
If we let $S(n)$ denote the space we use for solving our problem with length $n$ strings $s'$ and $t'$,
our space follows the recursion $S(n) = S(n/2) + \OO(n)$, with the base case that $S(\sqrt{n}) = \OO(n)$.
This sum is geometric and solves to $S(n) = \OO(n)$.
\end{proof}

\section{Open questions} \label{sec:open_questions}
% In this work, we extended the approach of Holm and T\v{e}tek \cite{HolmT23} in three ways.
% With our graph drawing algorithms, we obtain a recursive framework that works for the MPC model in the fully scalable regime where each machine only has a memory size $\cS = m^\delta$, for any given constant $\delta > 0$.  Using this framework, we obtain constant-round MPC algorithms for connected components, MST, and bipartition in embedded planar graphs.
% Incorporating the $\eps$-emulator of Chang, Krauthgamer, and Tan~\cite{ChangKT22} into our framework, we obtain constant-round $(1+\eps)$-approximation algorithms for SSSP and shortest cycle in embedded planar graphs. With our dual graph construction algorithm, we solve the maximum flow problem by reducing it to the  shortest cycle in the dual graph. 
 
%The computations of distances, cuts, and flows are difficult problems in the MPC model in that all existing works on this topic require a local memory of at least linear size: $\cS = \Omega(n)$~\cite{becker2021near,henzinger2019deterministic,ghaffari2018congested,ghaffari2020massively}. Our work is the first one that solves these problems in the fully scalable regime.

We believe that the new techniques developed in this work will be relevant to future research on geometric problems in MPC, which is an area where many fundamental questions remain relatively unexplored.
We list some of these open questions:

\begin{description}
\item [1. EMST in higher dimensions.]
Our work implies that two-dimensional Euclidean MST can be solved in $O(1)$ rounds in the MPC model in the fully scalable regime. It is a challenging open question to extend this result to higher dimensions. For the case where the dimension is a constant $d = O(1)$, Andoni, Nikolov, Onak, and Yaroslavtsev~\cite{andoni2014parallel} designed an $O(1)$-round algorithm for $(1+\eps)$-approximation of Euclidean MST.
Very recently, it was shown in~\cite{cohen2022streaming} that $(1+\eps)$-approximation is also possible in the regime of $d = \omega(1)$. Specifically, they showed that for any constants $\epsilon \in (0,1)$ and $\delta \in (0,1)$, a $(1+\eps)$-approximation of Euclidean MST can be computed in $O(1)$ rounds with local memory size $\cS = O(n^\delta) \cdot d^{O(1)}$.
It is still unknown whether Euclidean MST can be solved \emph{exactly} in $O(1)$ rounds even for $d = 3$.

\item[2. Approximate diameter and radius.]
For a planar embedded graph $G$ with $n$ vertices and $m$ edges, we use $O(n^2)$ total space for computing a $(1+\eps)$-approximate diameter and radius in $O(1)$ rounds with our algorithm for APSP. On the other hand, with $O(n)$ total space, we can compute a $(2+\eps)$-approximate diameter and radius in $O(1)$ rounds with our algorithm for SSSP. 
Does there exist an $O(1)$-round algorithm for computing $(1+\eps)$-approximate diameter or radius using $O(n)$ total space?
In the sequential setting, it was shown that a $(1+\eps)$-approximate diameter can be computed in $O(n)$ time by a sequence of results~\cite{WeimannY16,Cabello19, ChanS19, ChangKT22}. 
However, all sequential results on this line require shortest path separators, rather than the cycle separators used for divisions, so it is not clear how we can use these results in our framework. %normal ones.
%To obtain an $O(1)$ round algorithm in the MPC model by adapting similar techniques, it seems we need some sort of \emph{shortest path divisions} and a way to compute them in the MPC model.

\item[3. Exact solutions to distance-based problems.]
For embedded planar graphs, all our algorithms for computing distances, cuts, and flows are $(1+\eps)$-approximation, due to the use of $\eps$-emulator. It is unknown whether it is possible to improve these results to obtain exact solutions, without any approximation errors. This is likely a difficult open problem, as all existing results on exact distance computation in parallel, distributed, and semi-streaming models require polynomial round complexity or pass complexity~\cite{chang2020streaming,censor2019fast}. For exact distance computation, what advantages do we obtain from restricting ourselves to the case of embedded planar graphs or to the case where the distances are given by the actual distances between the points in Euclidean space?

\item[4. Geometric intersection graphs.]
There exist extensions of the planar separator theorem to some types of \emph{geometric intersection graphs}, such as the \emph{unit-disc graphs}~\cite{deberg2020framework,deberg2022clique}. These extensions were used to design sequential algorithms that have matching conditional lower bounds~\cite{deberg2020framework}. Is it possible to utilize the techniques developed in~\cite{deberg2020framework,deberg2022clique} to obtain improved parallel and distributed algorithms for some classes of geometric intersection graphs?

\item[5. Planarity testing and graph drawing.] 
All our planar graph algorithms assume that an embedding of the input planar graph $G$ is given. 
What are the round complexities of planarity testing and drawing a planar graph? Essentially nothing is known about these problems in the MPC model.
In the related $\CONGEST$ model of distributed computing, it is known that a combinatorial embedding of a planar network can be computed in near-diameter rounds $O(D \log n)$~\cite{ghaffari2016planar}, and after a combinatorial embedding is given, a polyline planar drawing with at most three bends per edge can be computed in $O(D)$ rounds~\cite{Sederholm2022}.
It is unlikely that a straight-line drawing of a planar graph can be computed in $o(\log n)$ rounds in the MPC model with strongly sublinear memory per machine, since such an $o(\log n)$-round algorithm, combined with our algorithms in this paper, would refute the 1-vs-2-cycles conjecture.
 Nanongkai and Scquizzato~\cite{nanongkai2022equivalence} showed that if the planarity testing problem can be solved in $o(\log n)$ rounds, then the 1-vs-2 cycle conjecture would also be refuted. 
\end{description}

\section*{Acknowledgements}
We thank Sariel Har-Peled for suggesting the sublinear time algorithm for constructing cuttings, Hsien-Chih Chang for pointing out the application to edit distance, and Timothy M.~Chan, Jeff Erickson, and Jack Spalding-Jamieson for helpful discussions. 

\appendix

\section*{Appendix}

\section{Proof of sublinear time algorithm for constructing cuttings}
\label{sec:sublinear_cuttings_proof}
In this section, we prove \cref{thm:sublinear_cuttings}.
The algorithm and the underlying ideas presented in this section are standard 
and are derived from classical randomized techniques in computational geometry. 
The key idea is that a sample of a set well approximates the old set, so we can sample a small set of edges and run an efficient algorithm to compute an optimal cutting on the sample.

We assume that the reader is familiar with VC-dimension.
Consider a range space $(X, \cR)$ of VC-dimension $d$. In particular, we are interested in $X$ being line segments and $\cR$ being the set of trapezoidal ranges, which has $VC$ dimension $O(1)$. Let $S$ be a finite subset of $X$ and let $R$ be a subset of $S$. 
For a range $H\in \cR$, let $H\sqcap S$ denote the set of line segments of $S$ intersecting $H$.
Then we define the following two quantities for each range $H\in \cR$.
\[ 
\text{The measure of $H$, } m(H) = \frac{|H\sqcap S|}{|S|}. 
\qquad 
\text{The estimate of $H$ by $R$, } s_R(H) = \frac{|H\sqcap R|}{|R|} 
\]

\begin{definition}
A subset $R \subseteq S$ is an $\eps$-approximation if for each range $H\in \cR$:
    \[ |\mu(H) - s_R(H)| \le \eps \]
\end{definition}

While this is enough for proving a sublinear time algorithm for cuttings, we can improve the bounds slightly
by using a closely related idea of relative approximations.

\begin{definition}
A subset $R \subseteq S$ is a \emph{relative $(p,\eps)$-approximation} if for each range $H\in \cR$:
\begin{enumerate}
    \item  If $\mu(H) \ge p$, then $(1-\eps) \mu(H) \le s_R(H) \le (1+\eps) \mu(H)$.
    \item  If $\mu(H) < p$, then $s_R(H) \le (1+\eps) p$.
\end{enumerate}
\end{definition}

The following theorem shows that random samples of relatively small sizes are relative $(p, \eps)$-approximations.
\begin{theorem}[\cite{LiLS01, Har-PeledS11}]
A sample $R$ of size $O\left(\eps^{-2}p^{-1}(d\log p^{-1} + \log \delta^{-1}) \right)$ is a relative $(p, \eps)$-approximation with probability at least $1-\delta$.
\end{theorem}

This directly implies a sublinear time algorithm for cuttings.

\sublinearcuttings*

\begin{proof}
Let $\Delta$ denote the set of all trapezoids in the plane. Trapezoids have VC-dimension $O(1)$.
For a set $S$ of $n$ non-crossing line segments, a sample $R\subseteq S$ of size $m = O(r(\log r + \log \delta^{-1}))$ is a $(1/(2r), 1/2)$-relative approximation with probability at least $1-\delta$.
Applying \Cref{thm:cuttings} we can construct a $(1/(2r))$-cutting $\Xi$ of $R$ of size $O(r)$ in $O(m\log r)$ time.
Consider a trapezoid $\sigma \in \Xi$, it must intersect at most $O(\log r + \log \delta^{-1})$ edges of $R$ by the definition of a cutting, so $s_R(\sigma) \le 1/(2r)$. 
Now let us consider if $\mu(\sigma) \ge 1/(2r)$. Since $R$ is a $(1/(2r), 1/2)$-approximation of $(S, \Delta)$ and $\sigma \in \Delta$, we know that $1/2 \mu(\sigma) \le s_R(\sigma)$, so $\mu(\sigma) \le 2s_R(\sigma) \le 1/r$. Thus we conclude that $\Xi$ is a $(1/r)$-cutting of $S$ with probability at least $1-\delta$.
\end{proof}

\paragraph{Remarks} Observe that the proof only relied on the fact that for line segments, ranges that are vertical trapezoids have $O(1)$ VC-dimension. We can easily replace line segments with bounded degree algebraic curves, or other pseudo-curves provided we modify our ranges to be the corresponding pseudo-trapezoids.
The only property of relative approximations we used was that for $\mu(h) \ge 1/(2r)$, $\mu(H) \le 2s_R(H)$. If we replaced the relative approximation with a $(1/(2r))$-approximation, the above property would still hold.
However, the bounds we would get would be worse.

\section{Deterministic cuttings in the MPC model}\label{sec:deterministic_cuttings}

As mentioned in the remark of \Cref{sec:sublinear_cuttings_proof}, we can use $\eps$-approximations instead.
We use the following two properties of $\eps$-approximations that have been observed many times in the literature that relative approximations do not have, that allow for the use of the \emph{merge-and-sketch} paradigm.
\begin{observation}[Lemma 5.33 of \cite{har2011}]
Let $(X, \cR)$ be a range space with VC-dimension $d$.
Then if $A_1$ is an $\eps$-approximation for $(X, \cR)$ and 
$A_2$ is an $\eps'$-approximation for $(A_1, \cR)$,
then $A_2$ is an $(\eps+\eps')$-approximation for $(X, \cR)$.
\end{observation}
\begin{observation}[Lemma 5.35 of \cite{har2011}]
Let $(X_1, \cR)$ and $(X_2, \cR)$ be two range spaces with VC-dimension $d$,
$A_1$ is an $\eps$-approximation for $(X_1, \cR)$ and 
$A_2$ is an $\eps$-approximation for $(X_2, \cR)$.
If $|A_1| = |A_2|$ and $|X_1| = |X_2|$, 
then $A_1 \cup A_2$ is an $\eps$-approximation for $(X_1\cup X_2, \cR)$.
\end{observation}

We also need deterministic computation of $\eps$-approximations.
We note that the space usage was never explicitly stated, but with 
careful analysis of the algorithm of Chazelle and Matou\v{s}ek \cite{ChazelleM96},
one can show it runs in linear space.
\begin{theorem}[\cite{Matousek95,ChazelleM96,Chazelle01}]
Let $(X, \cR)$ be a range space with VC-dimension $d$ for constant $d$, where $|X| = n$. 
Then one can compute an $\eps$-approximation for $(X, \cR)$ of size $O(\eps^{-2} \log(\eps^{-1}))$ 
deterministically in $O(n)$ space and $O(n\eps^{-2d}\log^d(\eps^{-1}))$ time.
\end{theorem}

This is all we need to prove our result on deterministic cuttings in the MPC model.
This is almost identical to the proof of \cite{ChazelleM96} 
for deterministic $\eps$-approximations in parallel,
albeit with a larger branching factor and different parameters
to ensure we finish in $O(1)$ rounds.
\thmdetcuttings*
\begin{proof}
Instead of computing a random sample, each machine can compute an $\eps$-approximation deterministically with $\eps = 1/(Cr)$ for a sufficiently large constant $C$.
The $\eps$-approximation will have size $O(r^2 \log r)$.
Ideally, we would send all of this to one machine, but that would require too much space. 
Instead, we will \emph{merge} as many of these approximations onto one machine, then create a \emph{sketch} by taking another $\eps$-approximation of the merged approximations. 

In one round, we can merge $O(\cS/r^3)$ of the $\eps$-approximations together onto one machine
since they would have total size $O(\cS/r^3)\cdot O(r^2\log r) = o(\cS)$.
We can do this in parallel amongst all the machines.
Afterward, we will take a sketch of the edges. To be precise, we compute another $\eps$-approximation of the approximations that is $O(r^2\log r)$ sized, and we can repeat until we are left with everything on one machine is the sum of all the $\eps$ we incurred from the merge steps.

This process terminates with a $O(r^2\log r)$ sized $\eps$-approximation on a single machine in $\log_{\cS/r^3} (n/\cS) = O(1/(\delta-3\alpha)) = O(1)$ rounds.
The total error accumulates at each sketch we performed. 
Since we chose $\eps = 1/(Cr)$, choosing $C$ to be sufficiently large, our final set of edges can be made to be a $(1/(2r))$-approximation of $S$.
At this point, we can apply the deterministic version of \Cref{thm:cuttings} as in \Cref{thm:sublinear_cuttings} on this approximation to get a $(1/r)$-cutting. 
\end{proof}

\bibliographystyle{alphaurl}
\bibliography{ref}